\documentclass[12pt, draftclsnofoot, journal, onecolumn]{IEEEtran}
\usepackage{blindtext}
\usepackage[T1]{fontenc}
\usepackage{textcomp}
\usepackage{amsmath}
\usepackage{amssymb}
\usepackage{bm}
\usepackage{mdwmath}
\usepackage{cases}
\usepackage{graphicx}
\usepackage{xcolor}
\definecolor{myblue}{rgb}{0,0.4980,1} 
\definecolor{myred}{rgb}{0.8706,0.1608,0.0627} 
\usepackage[caption=false,font=footnotesize,subrefformat=parens]{subfig}
\usepackage{makecell}
\usepackage{multirow}
\usepackage{acronym}
\usepackage[nosort,noadjust]{cite}
\usepackage{url}

\usepackage{hyperref}
\hypersetup{%
	bookmarksopen=true,%
	bookmarksnumbered=true,%
	pdfpagemode={UseOutlines},
	pdfstartview={FitH},%
	colorlinks=true,%
	linkcolor={myred},%
	citecolor={cyan}}

\usepackage{algpseudocode}
\newcounter{MYalgorithmic}
\renewcommand{\theMYalgorithmic}{\arabic{MYalgorithmic}}
\newcommand{\algcaption}[1]{%
	\refstepcounter{MYalgorithmic}%
	\textbf{Algorithm}~\textbf{\theMYalgorithmic}.~#1}
\newenvironment{MYalgorithmic}[5]
{
	\hrule height 1.2pt
	\vspace{3pt}
	#1{#2}%
	#3{#4}
	\vspace{3pt}
	\hrule height 0.5pt
	\vspace{3pt}
	#5
}
{
	\vspace{3pt}
	\hrule height 0.5pt
}

\newcounter{MYitem}[MYalgorithmic]
\renewcommand{\theMYitem}{\arabic{MYitem}}
\newcommand{\algitem}{%
	\refstepcounter{MYitem}%
	\textbf{\theMYitem)}}

\makeatletter
\newcommand{\MYlabel}[1]{\def\@currentlabel{\theALG@line}\label{#1}}
\makeatother

\newtheorem{problem}{\textbf{Problem}}
\newtheorem{lemma}{\textbf{Lemma}}
\newtheorem{theorem}{\textbf{Theorem}}
\newtheorem{remark}{\textbf{Remark}}
\newtheorem{corollary}{\textbf{Corollary}}
\newtheorem{subproblem}{\textbf{Sub-Problem}}

\newcommand{\upperroman}[1]{\uppercase\expandafter{\romannumeral#1}}

\newcommand{\myvec}[1]{\bm{\mathrm{#1}}}
\newcommand{\myunit}[1]{$ \mathrm{#1} $}
\newcommand{\myexp}{\mathrm{e}}
\DeclareMathOperator{\sbjto}{s.t.}
\DeclareMathOperator{\diag}{diag}

\newcommand{\MYnewpage}{%
	\ifCLASSOPTIONonecolumn%
		\ifCLASSOPTIONjournal%
			\typeout{The onecolumn journal mode.}%
			\newpage%
		\fi%
	\fi}
\begin{document}
\ifCLASSOPTIONonecolumn
	\typeout{The onecolumn mode.}
	\title{\LARGE Twin-Timescale Radio Resource Management for Ultra-Reliable and Low-Latency Vehicular Networks}
	\author{Haojun~Yang,~\IEEEmembership{Student Member,~IEEE},
		Kan~Zheng,~\IEEEmembership{Senior Member,~IEEE},
		Long~Zhao,~\IEEEmembership{Member,~IEEE},
		and~Lajos~Hanzo,~\IEEEmembership{Fellow,~IEEE}
		\thanks{Haojun~Yang, Kan~Zheng and Long~Zhao are with the Intelligent Computing and Communications ($ \text{IC}^\text{2} $) Lab, Wireless Signal Processing and Networks (WSPN) Lab, Key Laboratory of Universal Wireless Communications, Ministry of Education, Beijing University of Posts and Telecommunications (BUPT), Beijing, 100876, China (E-mail: \textsf{yanghaojun.yhj@bupt.edu.cn; zkan@bupt.edu.cn; z-long@bupt.edu.cn}).}
		\thanks{Lajos Hanzo is with the School of Electronics and Computer Science, University of Southampton, Southampton SO17 1BJ, U.K. (E-mail: \textsf{lh@ecs.soton.ac.uk}).}
	}
\else
	\typeout{The twocolumn mode.}
	\title{Twin-Timescale Radio Resource Management for Ultra-Reliable and Low-Latency Vehicular Networks}
	\author{Haojun~Yang,~\IEEEmembership{Student Member,~IEEE},
		Kan~Zheng,~\IEEEmembership{Senior Member,~IEEE},
		Long~Zhao,~\IEEEmembership{Member,~IEEE},
		and~Lajos~Hanzo,~\IEEEmembership{Fellow,~IEEE}
		\thanks{Haojun~Yang, Kan~Zheng and Long~Zhao are with the Intelligent Computing and Communications ($ \text{IC}^\text{2} $) Lab, Wireless Signal Processing and Networks (WSPN) Lab, Key Laboratory of Universal Wireless Communications, Ministry of Education, Beijing University of Posts and Telecommunications (BUPT), Beijing, 100876, China (E-mail: \textsf{yanghaojun.yhj@bupt.edu.cn; zkan@bupt.edu.cn; z-long@bupt.edu.cn}).}
		\thanks{Lajos Hanzo is with the School of Electronics and Computer Science, University of Southampton, Southampton SO17 1BJ, U.K. (E-mail: \textsf{lh@ecs.soton.ac.uk}).}
	}
\fi

\ifCLASSOPTIONonecolumn
	\typeout{The onecolumn mode.}
\else
	\typeout{The twocolumn mode.}
	\markboth{IEEE Transactions on Vehicular Technology}{Yang \MakeLowercase{\textit{et al.}}: Title}
\fi

\maketitle

\ifCLASSOPTIONonecolumn
	\typeout{The onecolumn mode.}
	\vspace*{-50pt}
\else
	\typeout{The twocolumn mode.}
\fi
\begin{abstract}
To efficiently support safety-related vehicular applications, the ultra-reliable and low-latency communication (URLLC) concept has become an indispensable component of vehicular networks (VNETs). Due to the high mobility of VNETs, exchanging near-instantaneous channel state information (CSI) and making reliable resource allocation decisions based on such short-term CSI evaluations are not practical. In this paper, we consider the downlink of a vehicle-to-infrastructure (V2I) system conceived for URLLC based on idealized perfect and realistic imperfect CSI. By exploiting the benefits of the massive MIMO concept, a two-stage radio resource allocation problem is formulated based on a novel twin-timescale perspective for avoiding the frequent exchange of near-instantaneous CSI. Specifically, based on the prevalent road-traffic density, Stage 1 is constructed for minimizing the worst-case transmission latency on a long-term timescale. In Stage 2, the base station allocates the total power at a short-term timescale according to the large-scale fading CSI encountered for minimizing the maximum transmission latency across all vehicular users. Then, a primary algorithm and a secondary algorithm are conceived for our V2I URLLC system to find the optimal solution of the twin-timescale resource allocation problem, with special emphasis on the complexity imposed. Finally, our simulation results show that the proposed resource allocation scheme significantly reduces the maximum transmission latency, and it is not sensitive to the fluctuation of road-traffic density.
\end{abstract}

\ifCLASSOPTIONonecolumn
	\typeout{The onecolumn mode.}
	\vspace*{-10pt}
\else
	\typeout{The twocolumn mode.}
\fi

\IEEEpeerreviewmaketitle

\MYnewpage


\section{Introduction}
\label{sec:Introduction}

\IEEEPARstart{G}{iven} the rapid development of the Internet of things (IoT), the design of radio access networks is undergoing a paradigm shift from promoting spectral-efficiency and energy-efficiency toward connecting everything~\cite{Qiu2018}. As a representative application scenario of IoT, vehicular networks (VNETs) support various vehicular applications for reducing the probability of traffic accidents and for improving both the driving and infotainment experience~\cite{zhengsurvey2015,MacHardy2018}. Traditional safety-related vehicular applications, such as platooning and road safety generally require a transmission latency within a few milliseconds and a reliability in terms of error probability down to $ 10^{-5} $ (or $ 10^{-6} $). Furthermore, in order to achieve the ultimate goal of autonomous driving on the road, future autonomous vehicles need sophisticated sensors to support safety-related applications. Ultra-reliable and low-latency communications (URLLCs) exchanging near-real-time information are also required to offer assistance to control systems~\cite{adv2015,Wang2019}. To this end, it is paramount to conceive URLLC techniques for future VNETs. In the conception of URLLC techniques, radio resource management plays a key role in improving the performance of VNETs, hence ground-breaking research is required.

The latency encountered may be classified according to the hierarchical architecture of networks. The physical (PHY) layer aims for reducing the transmission latency, while the media access control (MAC) layer aspires to reduce the queueing latency. The theoretical analysis of PHY layer radio resource management conceived for device-to-device-based vehicle-to-vehicle communications can be found in~\cite{Sun2016}, where both the transmission latency and reliability are considered simultaneously. However, the study of vehicular mobility was not considered in~\cite{Sun2016}. Moreover, neither the ergodic capacity nor the outage capacity are suitable for characterizing the tradeoff among the achievable rate, transmission latency and reliability in URLLCs~\cite{Polyanskiy2010,Hayashi2009,Giuseppe2016}. Hence, the recent advances in studying an attractive tradeoff among the rate, latency and reliability of multiple-antenna aided channels are illustrated in~\cite{Yang2014} utilizing finite blocklength theory. However, there is still a paucity of literature on addressing the URLLC problems of the PHY layer in VNETs under the consideration of vehicular mobility models.

With respect to the MAC layer, sophisticated techniques have been used for analyzing and optimizing the queueing latency, such as Markov decision processes~\cite{Zheng2016,Zheng2015}, stochastic network calculus~\cite{Yang2018,Forssell2019} and queueing theory~\cite{Choi2019}. Based on the Markov decision process framework, a minimum queueing delay-based radio resource allocation scheme is proposed in~\cite{Yang2017} for vehicle-to-vehicle communications, where the successful packet reception ratio is considered as the reliability metric. However, it is not practical to inform the base station of the global channel state information (CSI) including the links which are not connected to the base station. Moreover, the analysis in~\cite{Yang2017} has not been provided for URLLCs, since it was not based on radical advances in finite blocklength theory. Relying on finite blocklength theory and stochastic network calculus, a novel probabilistic delay bound is conceived in~\cite{Sebastian2015} for low-latency machine-to-machine applications. Furthermore, a URLLC cross-layer optimization framework is proposed in~\cite{She2018} by jointly considering the transmission latency and queueing latency. However, the advances mentioned above are not applicable to VNETs due to the lack of vehicular mobility model.

In addition to latency, reliability is another important performance metric for VNETs. In the MAC layer, reliability is generally characterized either by the stability of the queue~\cite{Cui2012} or by the probability of the queueing latency exceeding the tolerance threshold~\cite{Mei2018,Choi2019}. As for the PHY layer, typically the latency vs reliability (packet error rate) tradeoff is quantified~\cite{Johansson2015}, where one of the pivotal parameters of ensuring a high reliability is deemed to be the multi-antenna diversity gain. Hence, large-scale MIMO schemes constitute a promising solution of reducing latency and enhancing reliability in VNETs. On a similar note, massive MIMO schemes can also be adopted for formulating large-scale fading-based optimization problems by exploiting the channel hardening phenomenon~\cite{Marzetta2010,zhengsurvey2015MIMO,Zhao2019}.

Against this background, in this paper we conceive radio resource management for URLLC VNETs based on a realistic vehicular mobility model. To conceive URLLC VNETs, the following pair of crucial characteristics has to be considered:
\begin{enumerate}
\item The near-instantaneous CSI rapidly becomes outdated and its more frequent update imposes a high overhead, hence reducing the communication efficiency; and
\item Compared to the time-scale of CSI fluctuation on the millisecond level, that of the road-traffic fluctuation is higher (second level). Requiring resource allocation updates on the order of milliseconds would impose an excessive implementation complexity.
\end{enumerate}
In tackling the above-mentioned challenges in the downlink of a massive MIMO vehicle-to-infrastructure (V2I) system conceived for URLLC, our contributions are:
\begin{itemize}
\item We optimize the transmission latency based on a twin-timescale perspective for reducing the signaling overhead, where a macroscopic road-traffic model is adopted for charactering the relationship between the vehicular velocity and road-traffic density.

\item We first derive the transmission latency based on finite blocklength theory for the matched filter (MF) precoder and zero-forcing (ZF) precoder both under perfect and imperfect CSI.

\item Then, a two-stage radio resource allocation problem is formulated for our V2I URLLC system. In particular, based on the long-term timescale of road-traffic density, Stage 1 aims for optimizing the worst-case transmission latency by appropriately setting the system's bandwidth. For Stage 2, the base station (BS) allocates the total power based on the large-scale fading CSI for minimizing the maximum transmission latency guaranteeing fairness among all vehicular users (VUEs). Finally, the proposed primary and secondary radio resource allocation algorithms are invoked for optimally solving the above two problems at a reasonable complexity.

\item Our simulation results show that the proposed resource allocation scheme is capable of significantly reducing the maximum transmission latency, while flexibly accommodating the road-traffic fluctuation. Hence, the radio resources can be allocated at a larger time interval, which means the proposed algorithms can get the effect of semi-persistent scheduling. Additionally, the ZF precoder is shown to constitute a compelling choice for our V2I URLLC system.
\end{itemize}

The remainder of this paper is organized as follows. First of all, finite blocklength theory is introduced in Section~\ref{sec:Preliminary}. Section~\ref{sec:Model} describes the V2I URLLC system model, while Section~\ref{sec:Problem} formulates the twin-timescale resource allocation problem. Then, the proposed resource allocation algorithms are discussed in Section~\ref{sec:Solution}. Finally, Section~\ref{sec:Simulation} illustrates the simulation results, while our conclusions are offered in Section~\ref{sec:Conclusion}.

\textit{Notations:} Uppercase boldface letters and lowercase boldface letters denote matrices and vectors, respectively, while $ \diag\lbrace \myvec{a} \rbrace $ is a diagonal square matrix whose main diagonal is formed by vector $ \myvec{a} $. Furthermore, $ (\cdot)^\text{T} $, $ (\cdot)^\text{*} $, $ (\cdot)^\text{H} $ and $ (\cdot)^{-1} $ represent the transpose, conjugate, conjugate transpose and pseudo-inverse of a matrix/vector, respectively, while $ \left\| \cdot \right\| $ denotes the Euclidian norm, and $ [ \cdot ]_{kk} $ denotes the $ k $-th diagonal element of a square matrix. Finally, $ \mathbb{E}(\cdot) $ represents the mathematical expectation, while $ \mathbb{CN}(\mu,\sigma^2) $ is the complex Gaussian distribution with mean $ \mu $ and real/imaginary component variance $ \sigma^2/2 $.

\section{Preliminaries: Finite Blocklength Theory}
\label{sec:Preliminary}

The Shannon formula quantifies the error-free capacity at which information can be transmitted over a band-limited channel in the presence of noise and interferences,
\begin{align}
\label{Shannon}
C(\gamma)=\mathbb{E}\left[ B\log_2(1+\gamma) \right],
\end{align}
where $ B $ is the bandwidth and $ \gamma $ is the signal-to-interference-plus-noise ratio (SINR). This capacity can only be approached at the cost of excessive coding latency and complexity, i.e.,
\begin{align}
C(\gamma)=\lim_{\epsilon \rightarrow 0} C_{\epsilon}(\gamma,\epsilon)=\lim_{\epsilon \rightarrow 0}\lim_{n \rightarrow \infty} R(\gamma,n,\epsilon),
\end{align}
where $ C_{\epsilon}(\gamma,\epsilon) $ is the so-called outage capacity at the error probability $ \epsilon $. For the operational wireless systems, both the ergodic and outage capacity are reasonable performance metrics, because the packet size is typically large.

However, the assumption of large packet size does not meet the requirements of URLLCs. Thus, a more refined analysis of $ R(\gamma,n,\epsilon) $ is needed. Fortunately, during the last few years, significant progress has been made for satisfactorily addressing the problem of approximating $ R(\gamma,n,\epsilon) $~\cite{Polyanskiy2010,Hayashi2009,She2017}:
\begin{align}
\label{approx}
R(\gamma,L,\epsilon) \approx \mathbb{E}\left\lbrace B\left[ \log_2(1+\gamma)-\sqrt{\dfrac{V}{LB}}Q^{-1}(\epsilon) \right] \right\rbrace,
\end{align}
where $ Q^{-1}(\cdot) $ denotes the inverse of the Gaussian $ Q $-function and $ V $ is the so-called channel dispersion. For a complex channel, the channel dispersion is given by~\cite{Polyanskiy2010,Hayashi2009}:
\begin{align}
V = \left( 1- \dfrac{1}{(1+\gamma)^2} \right) \left( \log_2 \myexp \right)^2.
\end{align}
Furthermore, $ L $ is the transmission latency\footnote{In general, the one-way latency of the PHY layer entails both the propagation latency of electromagnetic wave and the coding latency. For example, for a $ 300 $~\myunit{m} cell radius, the propagation delay of $ 1 $~\textmu\myunit{s} is typically negligible. Hence, the coding latency dominates the one-way delay of the PHY layer's transmission latency.}, while $ LB $, which is also referred to as the blocklength of channel coding, represents the number of transmitted symbols. When $ LB $ is high enough, the approximation~\eqref{approx} approaches the ergodic capacity. Compared to the ergodic capacity, the approximation~\eqref{approx} also implies that the rate reduction is proportional to $ 1/\sqrt{LB} $, when aiming for meeting a specific error probability at a given packet size. Finite blocklength theory constitutes a powerful technique of dealing with the URLLC-related optimization problems. In addition to the contributions mentioned above, some new advances based on \eqref{approx} analyze and optimize the URLLC performance of 5G and IoT networks~\cite{Hu2019,Sun2019,Zhou2019}.

\section{V2I URLLC System Model}
\label{sec:Model}

\begin{figure}[!t]
	\centering
	\includegraphics[scale=0.5]{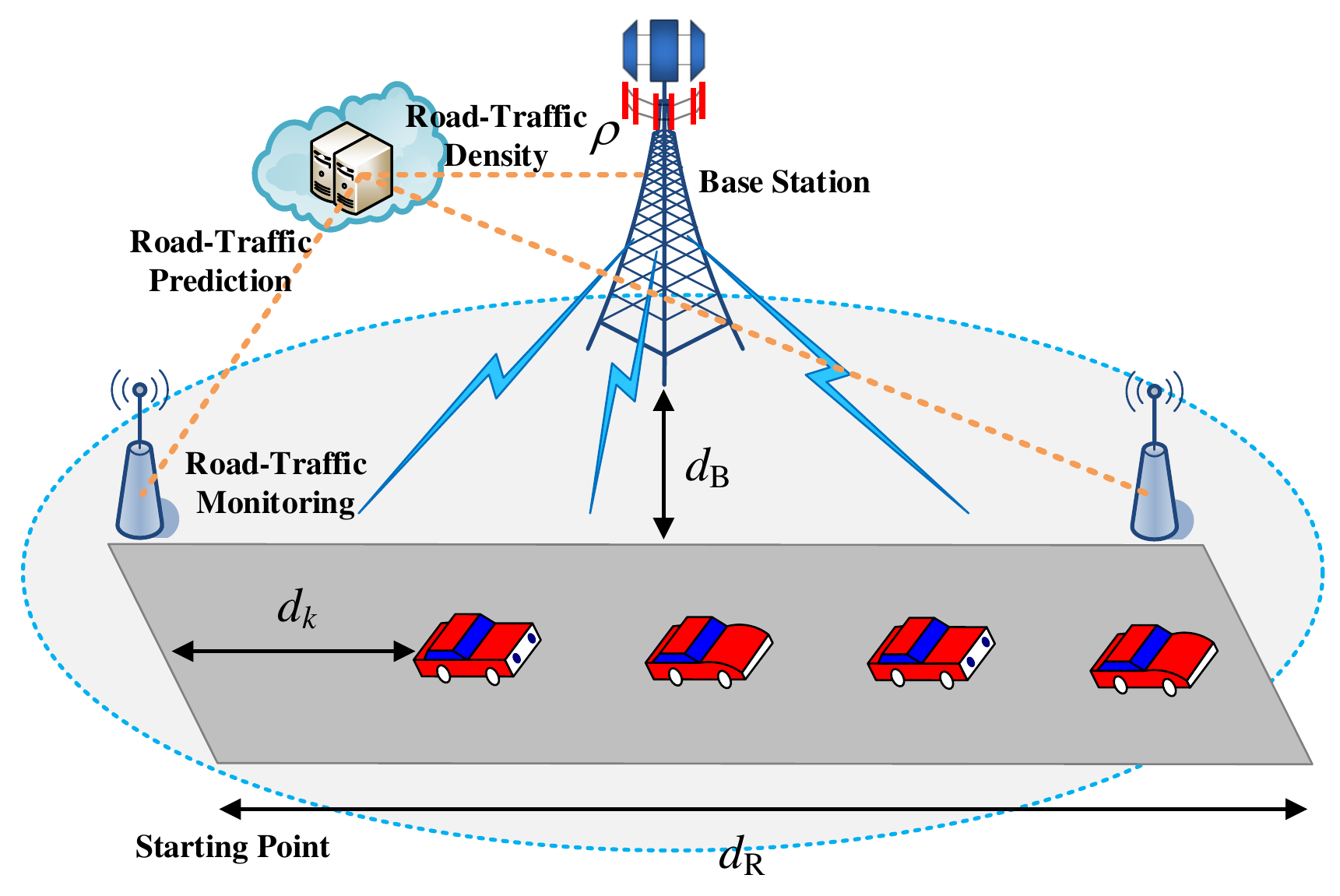}
	\caption{Illustration of system model.}
	\label{fig_Model}
\end{figure}

As shown in Fig.~\ref{fig_Model}, for the downlink of the V2I URLLC system, we consider a single roadside BS and a highway road segment of length $ d_\text{R} $. The BS is $ d_\text{B} $ meters away from the road. Furthermore, the BS employs $ M $ antennas and simultaneously sends information to $ K $ single-antenna aided VUEs (massive MIMO with $ M \gg K $). The system operates in the time-division duplex (TDD) mode.

\subsection{Road-Traffic Model}
According to the classic theory, the road-traffic model is divided into macroscopic and microscopic scales. Specially, the macroscopic model describes the average behavior of a certain number of vehicles at specific locations and instances, treating the road-traffic similarly to fluid dynamics. By contrast, the microscopic model describes the specific behavior of each individual entity (such as vehicle or pedestrian), hence it is more sophisticated than the macroscopic model. The open-source simulation of urban mobility (SUMO) mainly focuses on generating the microscopic model, namely the trajectory of vehicles. However, our proposed algorithms depend both on the road-traffic density and on the location information of all VUEs at a certain moment. Based on the macroscopic model, the above two types of information can be predicted and periodically reported by the traffic center~\cite{Lv2015,Polson2017}. To this end, we adopt the one-dimensional macroscopic model in this paper, which is also known as the Lighthill-Whitham-Richards (LWR) model~\cite{trafficflow}. Given the road-traffic density $ \rho $ and the road-traffic velocity $ v $, the road-traffic flow rate (flux) $ f $ is generally given by
\begin{align}
f = \rho v.
\end{align}
Numerous models have been proposed for characterizing how the vehicular velocity depends on the road-traffic density. In this paper, the Underwood model is employed, and its speed-density function can be written as~\cite{trafficflow}:
\begin{align}
\label{speed}
v(\rho) = v_\text{F} \exp\left( \dfrac{-\rho}{\rho_\text{m}} \right),
\end{align}
where $ v_\text{F} $ is the free-flowing velocity and $ \rho_\text{m} $ is the maximum density. Moreover, the road-traffic density can be used for modeling the number of VUEs, according to $ K = \rho d_\text{R} $.

\subsection{Channel Model}
All channels experience independent flat block-fading, i.e., they remain constant during a a coherence block, but change independently from one block to another. Let $ \myvec{G} = \myvec{D}^{1/2}\myvec{H} = \left[ \myvec{g}_1, \cdots, \myvec{g}_K \right]^\text{T} \in \mathbb{C}^{K \times M} $ denote the system's CSI, where the diagonal matrix $ \myvec{D} = \diag\left\lbrace \beta_1, \cdots, \beta_K \right\rbrace \in \mathbb{R}^{K \times K} $ and $ \myvec{H} = \left[ \myvec{h}_1, \cdots, \myvec{h}_K \right]^\text{T} \in \mathbb{C}^{K \times M} $ represent the large-scale fading and fast fading. The channel vector spanning from the BS to the $ k $-th VUE is given by $ \myvec{g}_k = \beta_k^{1/2}\myvec{h}_k \in \mathbb{C}^{M \times 1} $. Furthermore, imperfect channel estimation is considered~\cite{Kashyap2016,Zhao2017,Zhang2018}, i.e., $ \myvec{g}_k = \beta_k^{1/2}\myvec{h}_k = \beta_k^{1/2}\sqrt{\chi_k}\hat{\myvec{h}}_k + \beta_k^{1/2}\sqrt{1-\chi_k}\myvec{e}_k $, where $ \hat{\myvec{h}}_k $, $ \myvec{e}_k $ and $ \chi_k \in \mathbb{R} $ are the estimate, error and estimation accuracy of $ \myvec{h}_k $, respectively. If $ \chi_k = 1 $, $ \myvec{g}_k $ corresponds to perfect channel estimation. Finally, all random variables are independent and identically distributed (i.i.d.) complex Gaussian random variables with mean 0 and variance 1, namely $ \hat{h}_{km}, e_{km} \mathop \sim\limits^{\text{i.i.d.}} \mathbb{CN}(0,1) $.

\subsection{Transceiver Model}

\subsubsection{Transmitter}
It is widely exploited that low-complexity linear transmit precoders are capable of attaining asymptotically optimal performance in massive MIMO. Therefore we consider the MF precoder and ZF precoder in this paper. Let $ \myvec{V} = \left[ \myvec{v}_1, \cdots, \myvec{v}_K \right] \in \mathbb{C}^{M \times K} $ and $ \myvec{v}_k \in \mathbb{C}^{M \times 1} $ denote the precoder matrix and vector, where
\begin{align}
\myvec{V}=
\begin{cases}
\hat{\myvec{G}}^\text{H}, &\text{for MF},\\
\hat{\myvec{G}}^\text{H}\left( \hat{\myvec{G}}\hat{\myvec{G}}^\text{H} \right)^{-1}, &\text{for ZF},
\end{cases}
\end{align}
and $ \hat{\myvec{G}} = \left[ \hat{\myvec{g}}_1, \cdots, \hat{\myvec{g}}_K \right]^\text{T} $ with $ \hat{\myvec{g}}_k = \beta_k^{1/2}\hat{\myvec{h}}_k $. The normalized form of $ \myvec{V} $ is given by $ \myvec{W} = \left[ \myvec{w}_1, \cdots, \myvec{w}_K \right] \in \mathbb{C}^{M \times K} $ with
\begin{align}
\myvec{w}_k = \dfrac{\myvec{v}_k}{\left\| \myvec{v}_k \right\|}, \forall k.
\end{align}
As a result, given the symbol vector $ \myvec{s}=\left[ s_1, \cdots, s_K \right]^\text{T} \in \mathbb{C}^{K \times 1} $ with $ \mathbb{E}\left[ \myvec{s}\myvec{s}^\text{H} \right]=\myvec{I}_K $, the transmitted signal vector of all VUEs can be written as
\begin{align}
\label{signal}
\myvec{x} = \myvec{W}\myvec{P}^{1/2}\myvec{s},
\end{align}
where $ \myvec{P}=\diag\left\lbrace \frac{\myvec{p}}{M} \right\rbrace=\frac{1}{M}\diag\left\lbrace p_1, \cdots, p_K \right\rbrace \in \mathbb{R}^{K \times K} $ is the power allocation matrix and the total power at the BS is $ P_\text{B} $.

\subsubsection{Receiver}
The received signal vector is given by
\begin{align}
\myvec{y} = \myvec{G}\myvec{x}+\myvec{n} = \myvec{G}\myvec{W}\myvec{P}^{1/2}\myvec{s}+\myvec{n},
\end{align}
where $ \myvec{n} = \left[ n_1, \cdots, n_K \right]^\text{T} \in \mathbb{C}^{K \times 1} $ is the i.i.d. additive white Gaussian noise with $ n_{k} \mathop \sim\limits^{\text{i.i.d.}} \mathbb{CN}(0,\sigma^2) $. The received signal of the $ k $-th VUE can be written as
\begin{align}
y_k &= \sqrt{\dfrac{p_k}{M}}\sqrt{\chi_k}\hat{\myvec{g}}_k^\text{T}\myvec{w}_k s_k+\sum_{i=1,i \neq k}^{K} \sqrt{\dfrac{p_i}{M}}\sqrt{\chi_k}\hat{\myvec{g}}_k^\text{T}\myvec{w}_i s_i \notag \\
&\quad + \sum_{j=1}^{K} \sqrt{\dfrac{p_j}{M}}\sqrt{1-\chi_k}\beta_k^{1/2}\myvec{e}_k^\text{T}\myvec{w}_j s_j + n_k, \forall k.
\end{align}
Hence, the SINR of the $ k $-th VUE is given by
\ifCLASSOPTIONonecolumn
	\typeout{The onecolumn mode.}
	\begin{align}
	\gamma_k &= \dfrac{{\dfrac{p_k}{M}}\chi_k\left| \hat{\myvec{g}}_k^\text{T}\myvec{w}_k \right|^2}{\sum\limits_{i=1,i \neq k}^{K} \dfrac{p_i}{M}\chi_k\left| \hat{\myvec{g}}_k^\text{T}\myvec{w}_i \right|^2+\dfrac{P_\text{B}}{M}\beta_k\left( 1-\chi_k \right) +\sigma^2} \notag \\
	&=
	\begin{cases}
	\dfrac{{\dfrac{p_k}{M}}\chi_k\left| \hat{\myvec{g}}_k^\text{T}\dfrac{\hat{\myvec{g}}_k^{*}}{\left\| \hat{\myvec{g}}_k^{*} \right\|} \right|^2}{\sum\limits_{\begin{subarray}{l} i = 1, \\ i \neq k \end{subarray}}^{K} \dfrac{p_i}{M}\chi_k\left| \hat{\myvec{g}}_k^\text{T}\dfrac{\hat{\myvec{g}}_i^{*}}{\left\| \hat{\myvec{g}}_i^{*} \right\|} \right|^2+ \dfrac{P_\text{B}}{M}\beta_k\left( 1-\chi_k \right) +\sigma^2}, & \text{for MF},\\
	\dfrac{1}{\left[ \left( \hat{\myvec{G}}\hat{\myvec{G}}^\text{H} \right)^{-1} \right]_{kk}} \dfrac{\dfrac{p_k}{M}}{\dfrac{P_\text{B}}{M}\beta_k\left( 1-\chi_k \right) + \sigma^2}, & \text{for ZF}.
	\end{cases}
	\end{align}
\else
	\typeout{The twocolumn mode.}
	\begin{align}
	\gamma_k &= \dfrac{{\dfrac{p_k}{M}}\chi_k\left| \hat{\myvec{g}}_k^\text{T}\myvec{w}_k \right|^2}{\sum\limits_{i=1,i \neq k}^{K} \dfrac{p_i}{M}\chi_k\left| \hat{\myvec{g}}_k^\text{T}\myvec{w}_i \right|^2+\dfrac{P_\text{B}}{M}\beta_k\left( 1-\chi_k \right) +\sigma^2} \notag \\
	&=
	\begin{cases}
	\frac{{\dfrac{p_k}{M}}\chi_k\left| \hat{\myvec{g}}_k^\text{T}\dfrac{\hat{\myvec{g}}_k^{*}}{\left\| \hat{\myvec{g}}_k^{*} \right\|} \right|^2}{\sum\limits_{\begin{subarray}{l} i = 1, \\ i \neq k \end{subarray}}^{K} \dfrac{p_i}{M}\chi_k\left| \hat{\myvec{g}}_k^\text{T}\dfrac{\hat{\myvec{g}}_i^{*}}{\left\| \hat{\myvec{g}}_i^{*} \right\|} \right|^2+ \dfrac{P_\text{B}}{M}\beta_k\left( 1-\chi_k \right) +\sigma^2}, & \text{for MF},\\
	\dfrac{1}{\left[ \left( \hat{\myvec{G}}\hat{\myvec{G}}^\text{H} \right)^{-1} \right]_{kk}} \dfrac{\dfrac{p_k}{M}}{\dfrac{P_\text{B}}{M}\beta_k\left( 1-\chi_k \right) + \sigma^2}, & \text{for ZF}.
	\end{cases}
	\end{align}	
\fi
Recall from $ \hat{\myvec{g}}_k = \beta_k^{1/2}\hat{\myvec{h}}_k $ that the instantaneous $ \gamma_k^{-1} $ can be rewritten as
\ifCLASSOPTIONonecolumn
	\typeout{The onecolumn mode.}
	\begin{align}
	\gamma_k^{-1} =
	\begin{cases}
	\sum\limits_{\begin{subarray}{l} i = 1, \\ i \neq k \end{subarray}}^{K} \dfrac{p_i}{p_k}\left| \dfrac{\hat{\myvec{h}}_k^\text{T}}{\left\| \hat{\myvec{h}}_k^\text{T} \right\|}\dfrac{\hat{\myvec{h}}_i^{*}}{\left\| \hat{\myvec{h}}_i^{*} \right\|} \right|^2+\dfrac{P_\text{B}\beta_k\left( 1-\chi_k \right)+M\sigma^2}{p_k\chi_k\beta_k\left\| \hat{\myvec{h}}_k \right\|^2}, & \text{for MF},\\
	\dfrac{P_\text{B}\beta_k\left( 1-\chi_k \right)+M\sigma^2}{p_k\chi_k\beta_k}\left[ \left( \hat{\myvec{H}}\hat{\myvec{H}}^\text{H} \right)^{-1} \right]_{kk}, & \text{for ZF}.
	\end{cases}
	\end{align}
\else
	\typeout{The twocolumn mode.}
	\begin{align}
	\gamma_k^{-1} =
	\begin{cases}
	\sum\limits_{\begin{subarray}{l} i = 1, \\ i \neq k \end{subarray}}^{K} \frac{p_i}{p_k}\left| \frac{\hat{\myvec{h}}_k^\text{T}}{\left\| \hat{\myvec{h}}_k^\text{T} \right\|}\frac{\hat{\myvec{h}}_i^{*}}{\left\| \hat{\myvec{h}}_i^{*} \right\|} \right|^2+\frac{P_\text{B}\beta_k\left( 1-\chi_k \right)+M\sigma^2}{p_k\chi_k\beta_k\left\| \hat{\myvec{h}}_k \right\|^2}, & \text{for MF},\\
	\dfrac{P_\text{B}\beta_k\left( 1-\chi_k \right)+M\sigma^2}{p_k\chi_k\beta_k}\left[ \left( \hat{\myvec{H}}\hat{\myvec{H}}^\text{H} \right)^{-1} \right]_{kk}, & \text{for ZF}.
	\end{cases}
	\end{align}
\fi
Based on~\cite{Ngo2013}, we have the following distributions, i.e.,
\begin{align}
\Omega_\text{B}^\text{MF} &\triangleq \left| \dfrac{\hat{\myvec{h}}_k^\text{T}}{\left\| \hat{\myvec{h}}_k^\text{T} \right\|}\dfrac{\hat{\myvec{h}}_i^{*}}{\left\| \hat{\myvec{h}}_i^{*} \right\|} \right|^2 \sim \text{Beta}\left( 1,M-1 \right), \forall i \neq k, \\
\Omega_\text{G}^\text{MF} &\triangleq \dfrac{1}{\left\| \hat{\myvec{h}}_k \right\|^2} \sim \text{Gamma}^{-1}\left( M,1 \right), \forall k, \\
\Omega_\text{G}^\text{ZF} &\triangleq \left[ \left( \hat{\myvec{H}}\hat{\myvec{H}}^\text{H} \right)^{-1} \right]_{kk} \sim \text{Gamma}^{-1}\left( M-K+1,1 \right), \forall k.
\end{align}
Then $ \gamma_k^{-1} $ can be finally expressed as
\ifCLASSOPTIONonecolumn
	\typeout{The onecolumn mode.}
	\begin{align}
	\label{SINR}
	\gamma_k^{-1} =
	\begin{cases}
	\dfrac{P_\text{B}-p_k}{p_k} \Omega_\text{B}^\text{MF} + \dfrac{P_\text{B}\beta_k\left( 1-\chi_k \right)+M\sigma^2}{p_k\chi_k\beta_k} \Omega_\text{G}^\text{MF}, & \text{for MF},\\
	\dfrac{P_\text{B}\beta_k\left( 1-\chi_k \right)+M\sigma^2}{p_k\chi_k\beta_k}\Omega_\text{G}^\text{ZF}, & \text{for ZF}.
	\end{cases}
	\end{align}
\else
	\typeout{The twocolumn mode.}
	\begin{align}
	\label{SINR}
	\gamma_k^{-1} =
	\begin{cases}
	\frac{P_\text{B}-p_k}{p_k} \Omega_\text{B}^\text{MF} + \frac{P_\text{B}\beta_k\left( 1-\chi_k \right)+M\sigma^2}{p_k\chi_k\beta_k} \Omega_\text{G}^\text{MF}, & \text{for MF},\\
	\dfrac{P_\text{B}\beta_k\left( 1-\chi_k \right)+M\sigma^2}{p_k\chi_k\beta_k}\Omega_\text{G}^\text{ZF}, & \text{for ZF}.
	\end{cases}
	\end{align}
\fi

\section{Problem Formulation of Twin-Timescale Radio Resource Management}
\label{sec:Problem}

\begin{figure}[!t]
	\centering
	\includegraphics[scale=0.5]{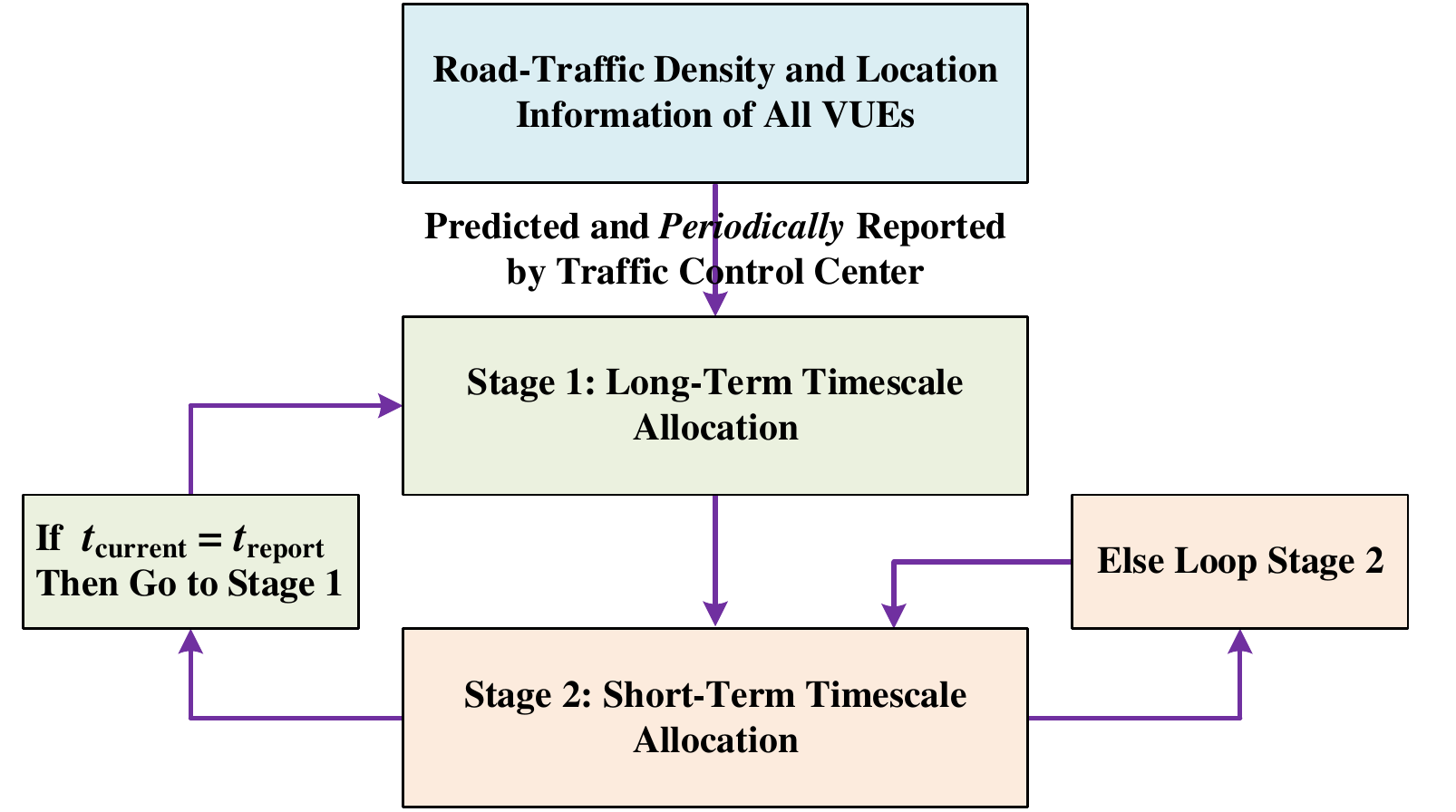}
	\caption{Illustration of twin-timescale radio resource management.}
	\label{fig_Timescale}
\end{figure}

In this treatise, we optimize the transmission latency of V2I communications based on the road-traffic density and location information, avoiding the frequent exchange of instantaneous global CSI. As the speed of mobility gradually increases, the fluctuation of the small-scale channel fading envelope generally becomes more rapid. However, the impact of small-scale fading can be mitigated by exploiting the channel hardening and asymptotic orthogonality of massive MIMO~\cite{Marzetta2010,Ngo2013,zhengsurvey2015MIMO}. As a result, the mobility of vehicles (the change of location information) only impacts the large-scale channel fading. Furthermore, compared to the time-scale of CSI fluctuation at the millisecond scale, that of the road-traffic (including the density and location information) fluctuation is on the scale of say 90~\myunit{km/h} (2.5~\myunit{cm/ms}). Therefore, considering the above facts, we formulate a twin-timescale large-scale fading-based radio resource allocation problem in order to reduce the signaling overhead imposed. As show in Fig.~\ref{fig_Timescale}, according to the road-traffic density, the first stage is constructed for optimizing the worst-case latency by setting the system's bandwidth from a long-term timescale. As for the second stage of the short-term timescale, the BS will allocate the total power based on the large-scale fading CSI (equivalent to location information) to minimize the maximum transmission latency for ensuring fairness amongst all VUEs.

\subsection{Large-Scale Fading-Based Transmission Latency}
In high-speed vehicular environments, small-scale channel fading tends to fluctuate rapidly, hence the instantaneous V2I CSI is often outdated. Furthermore, exchanging the instantaneous V2I CSI with the BS is not practical. As a result, large-scale fading-based allocation problems are generally considered by exploiting the channel hardening and asymptotic orthogonality of massive MIMO~\cite{Liu2018,Marzetta2010,Ngo2013,zhengsurvey2015MIMO}. The theorem for quantifying the maximum achievable rate is first formulated as follows.

\begin{theorem}
\label{theo1}	
When the MF or ZF precoder is employed in the downlink of our V2I URLLC system, the maximum achievable ergodic rate of the $ k $-th VUE can be approximated ($ M \gg 1 $) by~\cite{Polyanskiy2010,Hayashi2009,She2017}:
\begin{align}
\label{theorem1}
& R_k (\Gamma_k,L_k,\epsilon_k) = B\log_2\left( 1+\Gamma_k \right) \notag \\
&\quad \quad - \log_2 \myexp \sqrt{\dfrac{B}{L_k}}Q^{-1}(\epsilon_k)\sqrt{1-\left( 1+\Gamma_k \right)^{-2}}, \forall k,
\end{align}
where
\begin{align}
\label{Gamma}
\Gamma_k =
\begin{cases}
\dfrac{Mp_k}{P_\text{B}-p_k+\varphi_k}, &\text{for MF},\\
p_k\varphi_k, &\text{for ZF},
\end{cases}
\end{align}
and 
\begin{align}
\label{varphi}
\varphi_k =
\begin{cases}
\dfrac{P_\text{B}\beta_k\left( 1-\chi_k \right)+M\sigma^2}{\chi_k\beta_k}\dfrac{M}{M-1}, &\text{for MF},\\
\dfrac{\chi_k\beta_k\left( M-K \right)}{P_\text{B}\beta_k\left( 1-\chi_k \right)+M\sigma^2}, &\text{for ZF}.
\end{cases}
\end{align}
\end{theorem}

\begin{IEEEproof}
See Appendix~\ref{app:theo1}.
\end{IEEEproof}

According to Theorem~\ref{theo1}, we can obtain the following corollary for illustrating the transmission latency.

\begin{corollary}
\label{coro1}
Given the desired rate $ R_k (\Gamma_k,L_k,\epsilon_k) $ and the reliability $ \epsilon_k $, the transmission latency of the $ k $-th VUE can be expressed as
\begin{align}
\label{metric}
L_k = \left[ \dfrac{\sqrt{B}Q^{-1}(\epsilon_k)\log_2 \myexp \sqrt{1-\left( 1+\Gamma_k \right)^{-2}}}{B\log_2\left( 1+\Gamma_k \right)-R_k (\Gamma_k,L_k,\epsilon_k)} \right]^2, \forall k.
\end{align}
\end{corollary}

With respect to the reliability, $ Q^{-1}(\epsilon) $ with $ \epsilon \in [0,1] $ obeys the following properties,
\begin{align}
Q^{-1}(\epsilon)
\begin{cases}
<0, &\quad \epsilon > 0.5,\\
=0, &\quad \epsilon = 0.5,\\
>0, &\quad \epsilon < 0.5,
\end{cases}
\end{align}
and $ \lim_{\epsilon \rightarrow 0}Q^{-1}(\epsilon) = +\infty $. Thus, a higher reliability (also known as $ 1-\epsilon $) leads to a higher latency, and vice versa.

\subsection{Stage 1: Long-Term Timescale Allocation}
In somewhat simplistic, but plausible terms one could argue that doubling the system bandwidth allows doubling the bitrate, halving the framelength and hence halving the transmission latency. This plausible concept is also adopted by the 3GPP~\cite{3gpp38211} for reducing the transmission slot-duration. Another influential factor is the coding delay, because the ultimate Shannon capacity limit is only achievable, if the codeword-length tends to infinity. Broadly speaking, the system bandwidth can be appropriately adjusted for reducing the transmission latency, subject to the ubiquitous bandwidth constraints. Hence the system bandwidth determined in Stage 1 can also be viewed as part of the system design. Based on the long-term timescale of road-traffic density, the objective of Stage 1 is to optimize the worst-case transmission latency. Notice that the road-traffic density is valid for a limited road segment within about 50~\myunit{m} to 200~\myunit{m}~\cite{Lv2015,Polson2017}. Furthermore, our system model is based on a highway segment rather than on an urban scenario. The BS is deployed very close to the road segment, and thus generally there are no signal obstructions. To this end, the impact of shadow fading is ignored in this paper~\cite{Abbas2019,Park2019}. Without loss of generality, the worst situation is encountered in the scenario of assuming a largest distance from the BS in case of equal power allocation. The path loss at the largest distance from the BS is given by
\begin{align}
\beta_\text{W}=\theta\left[ d_\text{B}^2+\left( \dfrac{d_\text{R}}{2} \right)^2 \right]^{-\dfrac{\alpha}{2}},
\end{align}
where $ \theta $ is a constant related both to the antenna gain and to the carrier frequency, while $ \alpha $ is the path loss exponent. Then, with $ p_k = \frac{P_\text{B}}{K} = \frac{P_0 B}{\rho d_\text{R}} $, the worst-case SINR can be expressed as
\ifCLASSOPTIONonecolumn
	\typeout{The onecolumn mode.}
	\begin{align}
	\label{worstGam}
	\Gamma_\text{W} =
	\begin{cases}
	\dfrac{MP_0}{P_0\left( \rho d_\text{R}-1 \right)+\dfrac{P_0 \beta_\text{W}\left( 1-\chi_\text{th} \right)+M N_0}{\chi_\text{th}\beta_\text{W}}\dfrac{\rho d_\text{R}M}{M-1}}, & \text{for MF},\\
	\dfrac{P_0}{\rho d_\text{R}}\dfrac{\chi_\text{th}\beta_\text{W}\left( M-\rho d_\text{R} \right)}{P_0 \beta_\text{W}\left( 1-\chi_\text{th} \right)+M N_0}, & \text{for ZF},
	\end{cases}
	\end{align}
\else
	\typeout{The twocolumn mode.}
	\begin{align}
	\label{worstGam}
	\Gamma_\text{W} =
	\begin{cases}
	\dfrac{MP_0}{P_0\left( \rho d_\text{R}-1 \right)+\frac{P_0 \beta_\text{W}\left( 1-\chi_\text{th} \right)+M N_0}{\chi_\text{th}\beta_\text{W}}\frac{\rho d_\text{R}M}{M-1}}, & \text{for MF},\\
	\dfrac{P_0}{\rho d_\text{R}}\dfrac{\chi_\text{th}\beta_\text{W}\left( M-\rho d_\text{R} \right)}{P_0 \beta_\text{W}\left( 1-\chi_\text{th} \right)+M N_0}, & \text{for ZF},
	\end{cases}
	\end{align}
\fi
where $ P_0 $ and $ N_0 $ are the power spectral density of the transmitted signal and noise, respectively, while $ \chi_\text{th} $ is the minimum requested estimation accuracy, namely $ \chi_\text{th} = \min_{k}\left\lbrace \chi_k \right\rbrace $. Therefore, the worst-case transmission latency $ L_\text{W} $ is finally given by
\begin{align}
L_\text{W} = \left[ \dfrac{\sqrt{B}Q^{-1}(\epsilon_\text{W})\log_2 \myexp \sqrt{1-\left( 1+\Gamma_\text{W} \right)^{-2}}}{B\log_2\left( 1+\Gamma_\text{W} \right)-R_\text{W}} \right]^2,
\end{align}
where $ \epsilon_\text{W} = \min\limits_{k}\left\lbrace \epsilon_k \right\rbrace $ and $ R_\text{W} = \max\limits_{k}\left\lbrace R_k (\Gamma_k,L_k,\epsilon_k)  \right\rbrace $.

On the other hand, with \eqref{speed}, the maximum Doppler frequency is given by
\begin{align}
f_\text{MD} = \dfrac{v(\rho)}{\lambda} = \dfrac{f_\text{C}v_\text{F} \exp\left( -\rho/\rho_\text{m} \right)}{c},
\end{align}
where $ f_\text{C} $ is the carrier frequency, and $ c $ is the speed of light. Hence, the typical coherence time can be calculated as~\cite{Fleury1996}:
\begin{align}
T_\text{C} = \sqrt{\dfrac{9}{16\pi f_\text{MD}^2}} = \sqrt{\dfrac{9c^2}{16\pi f_\text{C}^2 v_\text{F}^2 \left[ \exp\left( -\rho/\rho_\text{m} \right) \right]^2}}.
\end{align}

The design of the existing wireless systems follows a rule that CSI remains constant during the coherence time~\cite{Dahlmanbook,3gpp38211}. Thus, motivated by this rule, we assume that the worst-case transmission latency is lower than the coherence time, which is similar with the design of the LTE-based systems. On the basis of the above derivations, the following optimization problem is formulated.

\begin{problem}[Long-Term Timescale Bandwidth Allocation]
\label{pro1}
Let $ \delta $ denote a very small threshold constant. Given the road-traffic density $ \rho $, the long-term timescale bandwidth allocation problem of Stage 1 is formulated as
\begin{align}
\label{problem1}
\textbf{\text{P\ref{pro1}:}}\quad  L_\text{W}(B,\rho)\leqslant \delta T_\text{C}(\rho).
\end{align}
Typically, $ \delta $ is equal to $ \frac{1}{20} $ for the LTE-based systems~\cite{Dahlmanbook}. Hence, in our simulations, we let $ \delta $ be equal to $ \frac{1}{20} $ or $ \frac{1}{40} $.
\end{problem}

\subsection{Stage 2: Short-Term Timescale Allocation}
By solving Problem~\ref{pro1}, the total power at the BS can be obtained, namely $ P_\text{B}=P_0B^*(\rho) $. In Stage 2, the BS allocates the total power based on the short-term timescale large-scale fading CSI of all VUEs for minimizing the maximum transmission latency. Then the following optimization problem is formulated.

\begin{problem}[Short-Term Timescale Power Allocation]
\label{pro2}
Given the desired rate $ R_k (\Gamma_k,L_k,\epsilon_k) $, reliability $ \epsilon_k $, total power $ P_\text{B} $ and the large-scale fading CSI of all VUEs, the optimization objective of Stage 2 is to minimize the maximum transmission latency of all VUEs ensuring fairness amongst all VUEs, i.e.,
\begin{subequations}
\label{problem2}
\begin{align}
\textbf{\text{P\ref{pro2}:}}\quad \min_{\myvec{p}}~ & \max_{k} \left\lbrace L_k \right\rbrace \\
\label{problem2con1}
\sbjto~ & \sum_{k=1}^{K}p_k = P_\text{B}, \\
\label{problem2con2}
& p_k\geqslant0,\forall k.
\end{align}
\end{subequations}
\end{problem}

\section{Twin-Timescale Resource Allocation Algorithms for V2I URLLC Scystem}
\label{sec:Solution}

In this section, the solutions to the above pair of problems are first studied, and then efficient twin-timescale allocation algorithms are proposed for our V2I URLLC system.

\subsection{Solution to Problem~\ref{pro1}}
According to Problem~\ref{pro1}, we have
\begin{align}
\dfrac{\sqrt{B}Q^{-1}(\epsilon_\text{W})\log_2 \myexp \sqrt{1-\left( 1+\Gamma_\text{W} \right)^{-2}}}{B\log_2\left( 1+\Gamma_\text{W} \right)-R_\text{W} (\gamma,L,\epsilon)}\leqslant \sqrt{\delta T_\text{C}},
\end{align}
which shows that
\begin{align}
& B\sqrt{\delta T_\text{C}}\log_2\left( 1+\Gamma_\text{W} \right)-\sqrt{\delta T_\text{C}}R_\text{W} (\gamma,L,\epsilon) \notag \\
&\quad -\sqrt{B}Q^{-1}(\epsilon_\text{W})\log_2 \myexp \sqrt{1-\left( 1+\Gamma_\text{W} \right)^{-2}} \geqslant 0.
\end{align}
Upon defining
\begin{align}
& \Delta = \left[ Q^{-1}(\epsilon_\text{W}) \right]^2 \log_2^2 e \left[ 1-\left( 1+\Gamma_\text{W} \right)^{-2} \right] \notag \\
&\quad + 4\delta T_\text{C}\log_2\left( 1+\Gamma_\text{W} \right)R_\text{W} (\gamma,L,\epsilon)>0,
\end{align}
the optimal system bandwidth $ B^* $ can be finally expressed as
\begin{align}
\label{optimalB}
B^* = \left[ \dfrac{Q^{-1}(\epsilon_\text{W})\log_2 \myexp \sqrt{1-\left( 1+\Gamma_\text{W} \right)^{-2}}+\sqrt{\Delta}}{2\sqrt{\delta T_\text{C}}\log_2\left( 1+\Gamma_\text{W} \right)} \right]^2,
\end{align}
where $ \Gamma_\text{W} $ can be found in \eqref{worstGam}. Given the system bandwidth, the total power at the BS is $ P_\text{B}=P_0B^* $.

\subsection{Solution to Problem~\ref{pro2}}

\subsubsection{Problem Transformation}
Problem~\ref{pro2} can be transformed into the following form:
\begin{subequations}
\label{problem2trans}
\begin{align}
\textbf{\text{P\ref{pro2}:}}\quad \max_{\myvec{p}}~ & \min_{k} \left\lbrace -\sqrt{L_k} \right\rbrace \\
\sbjto~ & \eqref{problem2con1}-\eqref{problem2con2}.
\end{align}
\end{subequations}

\subsubsection{Property of Objective Function}
Recall from \eqref{metric} that $ -\sqrt{L_k} $ can be expressed as
\begin{align}
\label{objective}
\dfrac{-\sqrt{B^*}Q^{-1}(\epsilon_k)\log_2 \myexp \sqrt{1-\left( 1+\Gamma_k \right)^{-2}}}{B^* \log_2\left( 1+\Gamma_k \right)-R_k (\Gamma_k,L_k,\epsilon_k)} \triangleq \dfrac{f_k(\myvec{p})}{g_k(\myvec{p})}, \forall k.
\end{align}

\begin{theorem}
\label{theo2}
When the MF or ZF precoder is employed in the downlink of our V2I URLLC system, $ -\sqrt{L_k} $ is a quasi-concave function of $ p_k $.
\end{theorem}

\begin{IEEEproof}
See Appendix~\ref{app:theo2}.
\end{IEEEproof}

Given the definition of quasi-concavity, $ \min_{k}\left\lbrace -\sqrt{L_k} \right\rbrace $ is also a quasi-concave function. By contrast, even if $ -\sqrt{L_k}, \forall k $ is pseudo-concave, $ \min_{k}\left\lbrace -\sqrt{L_k} \right\rbrace $ still cannot be pseudo-concave, since the function $ \min(\cdot) $ is not differentiable.

\subsubsection{Optimal Solution}
Since the objective function of \eqref{problem2trans} is QC, the Karush-Kuhn-Tucker (KKT) condition cannot be adopted. However, we will show that Problem~\ref{pro2} can still be solved optimally by an efficient algorithm.

Based on \eqref{objective}, the following auxiliary function $ F(\eta) $ is constructed,
\begin{align}
\label{auxiliary}
F(\eta) = \max_{\myvec{p}}\min_{k} \left\lbrace f_k(\myvec{p})-\eta g_k(\myvec{p}) \right\rbrace.
\end{align}
Intuitively, $ F(\eta) $ is strictly monotonically decreasing and $ F(\eta)=0 $ has a unique root.

\begin{lemma}
\label{lem1}
For any $ \tilde{\myvec{p}} \in \mathcal{P} $, we have $ F(\eta_{\tilde{\myvec{p}}}) \geqslant 0 $ with $ \eta_{\tilde{\myvec{p}}} = \min_{k}\left\lbrace f_k(\tilde{\myvec{p}})/g_k(\tilde{\myvec{p}}) \right\rbrace $, and the equality holds when $ \tilde{\myvec{p}} = \arg \max\limits_{\myvec{p}}\min\limits_{k} \left\lbrace f_k(\myvec{p})-\eta_{\tilde{\myvec{p}}} g_k(\myvec{p}) \right\rbrace $.
\end{lemma}

\begin{IEEEproof}
See Appendix~\ref{app:lem1}.
\end{IEEEproof}

Based on Lemma~\ref{lem1}, the following theorem describing the optimal allocation is formulated.

\begin{theorem}
\label{theo3}
For $ \myvec{p}^* \in \mathcal{P} $ and $ \eta^* = \min_{k}\left\lbrace f_k(\myvec{p}^*)/g_k(\myvec{p}^*) \right\rbrace $, $ \myvec{p}^* $ is the optimal solution of \eqref{problem2trans} if and only if
\begin{align}
\label{theorem3}
\myvec{p}^* = \arg \max\limits_{\myvec{p}}\min\limits_{k} \left\lbrace f_k(\myvec{p})-\eta^* g_k(\myvec{p}) \right\rbrace.
\end{align}
\end{theorem}

\begin{IEEEproof}
It relies on proving both sufficiency and necessity. Firstly, let $ \myvec{p}^* $ be the optimal power allocation of \eqref{problem2trans},
\begin{align}
\eta^* = \min_{k}\left\lbrace \dfrac{f_k(\myvec{p}^*)}{g_k(\myvec{p}^*)} \right\rbrace \geqslant \min_{k}\left\lbrace \dfrac{f_k(\myvec{p})}{g_k(\myvec{p})} \right\rbrace, \forall \myvec{p} \in \mathcal{P}.
\end{align}
Therefore,
\begin{align}
\min_{k}\left\lbrace f_k(\myvec{p}) - \eta^* g_k(\myvec{p}) \right\rbrace &\leqslant 0, \forall \myvec{p} \in \mathcal{P}, \\
\min_{k}\left\lbrace f_k(\myvec{p}^*) - \eta^* g_k(\myvec{p}^*) \right\rbrace &= 0,
\end{align}
which yields \eqref{theorem3} and $ F(\eta^*)=0 $.

Conversely, let $ \myvec{p}^* = \arg \max\limits_{\myvec{p}}\min\limits_{k} \left\lbrace f_k(\myvec{p})-\eta^* g_k(\myvec{p}) \right\rbrace $, i.e.,
\begin{align}
\label{necc}
\min_{k}\left\lbrace f_k(\myvec{p}) - \eta^* g_k(\myvec{p}) \right\rbrace &\leqslant \min_{k}\left\lbrace f_k(\myvec{p}^*) - \eta^* g_k(\myvec{p}^*) \right\rbrace \notag \\
&= F(\eta^*) \overset{\text{(a)}}{=} 0, \forall \myvec{p} \in \mathcal{P},
\end{align}
where (a) follows from Lemma~\ref{lem1}. Then, \eqref{necc} shows that
\begin{align}
\eta^* &\geqslant \min_{k}\left\lbrace \dfrac{f_k(\myvec{p})}{g_k(\myvec{p})} \right\rbrace, \forall \myvec{p} \in \mathcal{P}, \\
\eta^* &= \min_{k}\left\lbrace \dfrac{f_k(\myvec{p}^*)}{g_k(\myvec{p}^*)} \right\rbrace.
\end{align}
Thus $ \myvec{p}^* $ is the optimal power allocation.
\end{IEEEproof}

\begin{corollary}
\label{coro2}
Based on the monotonicity of $ F(\eta) $ and Theorem~\ref{theo3}, for any $ \eta $, we have
\begin{align}
\label{optimaleta}
F(\eta)
\begin{cases}
<0, &\quad \eta > \eta^*,\\
=0, &\quad \eta = \eta^*,\\
>0, &\quad \eta < \eta^*,
\end{cases}
\end{align}
where $ \eta^* $ is the optimal objective function value of \eqref{problem2trans}.
\end{corollary}

According to Theorem~\ref{theo3} and Corollary~\ref{coro2}, solving Problem~\ref{pro2} is equivalent to finding the unique root of $ F(\eta)=0 $.

\subsection{Twin-Timescale Resource Allocation Algorithms for V2I URLLC System}
By taking into account the above solutions, the following primary algorithm and secondary algorithm are proposed for the optimal V2I URLLC resource allocation.

\subsubsection{Primary Algorithm}

\begin{figure}[!t]
\begin{MYalgorithmic}
\algcaption{Road-traffic-aware twin-timescale resource allocation algorithm for V2I URLLC system}
\label{alg1}
\begin{algorithmic}[5]
\renewcommand{\algorithmicrequire}{\textbf{Initialization:}}
\Require
\State \labelitemi~Periodically reported road-traffic density $ \rho $.
\State \labelitemi~The location information of all VUEs.
\State \labelitemi~The rate requirement $ R_k (\Gamma_k,L_k,\epsilon_k) $ and reliability requirement $ \epsilon_k $ of each VUE.
\State \labelitemi~Iterative index $ j=0 $, maximum iterative tolerance $ \zeta_\text{P} > 0 $ and initial objective value $ \eta_0 < 0 $.
\renewcommand{\algorithmicrequire}{\textbf{Stage 1: Long-term timescale resource allocation}}
\Require
\State \algitem~Calculate the optimal system bandwidth $ B^* $ based on \eqref{optimalB}.
\State \algitem~Calculate the total power at the BS with $ P_\text{B}=P_0B^* $.
\renewcommand{\algorithmicrequire}{\textbf{Stage 2: Short-term timescale resource allocation}}
\Require
\Loop \quad $ (j) $
\State {\algitem\label{subalg}}~Based on Algorithm~\ref{alg2}, solve the power allocation $ \myvec{p}_j $ of Sub-problem~\ref{subpro1} with the given $ \eta_j $.
\State \algitem~Calculate $ F_j = \min_{k} \left\lbrace f_k(\myvec{p}_j)-\eta_j g_k(\myvec{p}_j) \right\rbrace $.
\If{$ F_j \leqslant \zeta_\text{P} $}
\State Update $ \myvec{p}^*=\myvec{p}_j $ and $ \eta^* = \min_{k}\left\lbrace \dfrac{f_k(\myvec{p}^*)}{g_k(\myvec{p}^*)} \right\rbrace $.
\State Set $ j=j+1 $ and \textbf{\textit{break}}.
\Else~~~\textbf{\textit{continue}}.
\EndIf
\State {\algitem\label{rule}}~Update $ \eta_{j+1} = \min_{k}\left\lbrace f_k(\myvec{p}_j)/g_k(\myvec{p}_j) \right\rbrace $.
\State \algitem~Set $ j=j+1 $.
\State \algitem~\textbf{If} the current slot is the moment of reporting periodic road-traffic density information, \textbf{Then} \textit{go to Stage 1} and \textbf{\textit{continue}}; \textbf{Else} \textit{loop Stage 2}.
\EndLoop
\end{algorithmic}
\end{MYalgorithmic}
\end{figure}

As shown in~\cite{Shen2018}, there are two classic techniques of solving fractional optimization problems, namely Charnes-Cooper transform and Dinkelbach's method. In simple terms, Dinkelbach's method can be interpreted as being reminiscent of Newton's method applied to the convex function $ F(\eta) $~\cite{Zappone2015}. Compared to Charnes-Cooper transform, Dinkelbach's method has the advantage of requiring no extra constraints~\cite{Shen2018}. Therefore, we opt for Dinkelbach's method in this paper. Based on the Dinkelbach's method~\cite{Zappone2015}, a twin-timescale iterative algorithm is put forward for jointly solving Problem~\ref{pro1} and \ref{pro2}, which is described in Algorithm~\ref{alg1}. The following theorem illustrates the convergence and optimality of Algorithm~\ref{alg1}.

\begin{theorem}
\label{theo4}
Algorithm~\ref{alg1} must converge to the optimal power allocation.
\end{theorem}

\begin{IEEEproof}
According to Step~\ref{rule} in Algorithm~\ref{alg1}, we have
\begin{align}
F_j &= \min_{k} \left\lbrace f_k(\myvec{p}_j)-\eta_j g_k(\myvec{p}_j) \right\rbrace \notag \\
&\overset{\text{(a)}}{=} \min_{k} \left\lbrace \left( \eta_{j+1}-\eta_j \right) g_k(\myvec{p}_j) \right\rbrace \notag \\
&\overset{\text{(b)}}{\geqslant} 0.
\end{align}
(a) follows from
\begin{align}
\dfrac{f_k(\myvec{p}_j)}{g_k(\myvec{p}_j)} \geqslant \eta_{j+1}, \forall k \Rightarrow f_k(\myvec{p}_j) \geqslant \eta_{j+1}g_k(\myvec{p}_j),
\end{align}
where the equality holds when $ k = \arg\min_{k}\left\lbrace f_k(\myvec{p}_j) / g_k(\myvec{p}_j) \right\rbrace $. (b) follows from Lemma~\ref{lem1}. This implies that when the algorithm does not achieve convergence, we have $ \eta_{j+1}>\eta_j $. Therefore, as the iterations proceed, $ \eta $ is gradually increased and $ F(\eta) $ is gradually decreased, hence Algorithm~\ref{alg1} converges.

The optimality can be proved by the method of contradiction. Let $ \eta_{\tilde{\myvec{p}}}=\min_{k}\left\lbrace f_k(\tilde{\myvec{p}})/g_k(\tilde{\myvec{p}}) \right\rbrace<\eta^* $ be an improved, but sub-optimal value. Given the monotonicity and convergence of $ F(\eta) $, we have $ F(\eta_{\tilde{\myvec{p}}})>F(\eta^*) $ and $ F(\eta_{\tilde{\myvec{p}}})=0 $. On the other hand, based on Theorem~\ref{theo3}, we have $ F(\eta^*)=0 $, which leads to a contradiction.
\end{IEEEproof}

\subsubsection{Secondary Algorithm}

\begin{figure}[!t]
\begin{MYalgorithmic}
\algcaption{Max-min resource allocation algorithm for Sub-problem~\ref{subpro1}}
\label{alg2}
\begin{algorithmic}[5]
\renewcommand{\algorithmicrequire}{\textbf{Initialization:}}
\Require
\State \labelitemi~The location information of all VUEs.
\State \labelitemi~The rate requirement $ R_k (\Gamma_k,L_k,\epsilon_k) $ and reliability requirement $ \epsilon_k $ of each VUE.
\State \labelitemi~Total power $ P_\text{B} $ and initial power allocation $ \myvec{p}_0=[p_1^0, \cdots, p_K^0]^\text{T} $ with $ p_k^0=P_\text{B}/K, \forall k $.
\State \labelitemi~Iterative index $ i=0 $, maximum iterative tolerance $ \zeta_\text{S} > 0 $ and adjustable allocation step length $ \mu_0 \in (0,P_\text{B}) $.
\renewcommand{\algorithmicrequire}{\textbf{Power allocation for Sub-problem~\ref{subpro1}:}}
\Require
\State \algitem~Calculate initial $ \myvec{H}_0=[h_1^0(\myvec{p}_0), \cdots, h_K^0(\myvec{p}_0)]^\text{T} $ with $ h_k^0(\myvec{p}_0) = f_k(\myvec{p}_0)-\eta_j g_k(\myvec{p}_0) $.
\Loop \quad $ (i) $
\State \algitem~Set $ i=i+1 $. 
\State \algitem~Calculate $ k_{i}^\text{m}=\arg\min_{k}\left\lbrace \myvec{H}_{i-1} \right\rbrace $ and $ k_{i}^\text{M}=\arg\max_{k}\left\lbrace \myvec{H}_{i-1} \right\rbrace $.
\State \algitem~Update $ \myvec{p}_i $ with $ p_{k_i^\text{m}}^{i} = p_{k_i^\text{m}}^{i-1} + \mu_{i-1} $, $ p_{k_i^\text{M}}^{i} = p_{k_i^\text{M}}^{i-1} - \mu_{i-1} $ and $ p_{k}^{i} = p_{k}^{i-1} (\forall k \neq k_i^\text{m},k_i^\text{M}) $.
\State \algitem~Calculate $ \myvec{H}_i $ with $ h_k^i(\myvec{p}_i) = f_k(\myvec{p}_i)-\eta_j g_k(\myvec{p}_i) $.
\If{$ \max_{k}\left\lbrace \myvec{H}_i \right\rbrace - \min_{k}\left\lbrace \myvec{H}_i \right\rbrace \leqslant \zeta_\text{S} $}
\State Update $ \myvec{p}_j=\myvec{p}_i $ and \textbf{\textit{break}}.
\Else
\If{$ h_{k_{i}^\text{m}}^{i}(\myvec{p}_{i}) \leqslant h_{k_{i}^\text{m}}^{i-1}(\myvec{p}_{i-1}) $ or $ h_{k_{i}^\text{M}}^{i}(\myvec{p}_{i}) \geqslant h_{k_{i}^\text{M}}^{i-1}(\myvec{p}_{i-1}) $ or $ p_{k_i^\text{M}}^{i} \leqslant 0 $ or $ \max_{k}\left\lbrace \myvec{H}_{i-1} \right\rbrace - \min_{k}\left\lbrace \myvec{H}_{i-1} \right\rbrace \leqslant \max_{k}\left\lbrace \myvec{H}_i \right\rbrace - \min_{k}\left\lbrace \myvec{H}_i \right\rbrace $}
\State Update $ \mu_i = \mu_{i-1}/2 $, $ \myvec{p}_i = \myvec{p}_{i-1} $ and $ \myvec{H}_i = \myvec{H}_{i-1} $.
\Else
\State Update $ \mu_i = \mu_{i-1} $ and \textbf{\textit{continue}}.
\EndIf
\EndIf
\EndLoop
\end{algorithmic}
\end{MYalgorithmic}
\end{figure}

As illustrated in Algorithm~\ref{alg1}, Step~\ref{subalg} raises the following optimization problem.

\begin{subproblem}[Max-Min Power Allocation]
\label{subpro1}
Given $ \eta_j < 0 $, the max-min power allocation problem is formulated as
\begin{subequations}
\label{subproblem1}
\begin{align}
\textbf{\text{SP\ref{subpro1}:}}\quad \max_{\myvec{p}}~ & \min_{k} \left\lbrace f_k(\myvec{p})-\eta_j g_k(\myvec{p}) \right\rbrace \\
\label{subproblem1con}
\sbjto~ & \eqref{problem2con1}-\eqref{problem2con2}.
\end{align}
\end{subequations}
\end{subproblem}

\ifCLASSOPTIONonecolumn
	\typeout{The onecolumn mode.}
	To solve Sub-problem~\ref{subpro1}, the monotonicity of $ h_k(\myvec{p}) = f_k(\myvec{p})-\eta_j g_k(\myvec{p}) $ is first studied. The proof of the ZF is similar to that of the MF, hence only the case of the MF is shown here. The derivative $ \partial h_k^\text{MF}(\myvec{p})/\partial p_k =\partial f_k^\text{MF}(\myvec{p})/\partial p_k - \eta_j \partial g_k^\text{MF}(\myvec{p})/\partial p_k $ is given by
	\begin{align}
	\label{diffh}
	\dfrac{\partial h_k^\text{MF}(\myvec{p})}{\partial p_k} = \dfrac{\sqrt{B^*}\left( \dfrac{\Gamma_k^\text{MF}}{p_k}+\dfrac{(\Gamma_k^\text{MF})^2}{Mp_k} \right)}{\sqrt{1-\left( 1+\Gamma_k^\text{MF} \right)^{-2}}\left( 1+\Gamma_k^\text{MF} \right)^3 \ln 2} \left[ -Q^{-1}(\epsilon_k)-\eta_j \sqrt{B^*}\sqrt{1-\left( 1+\Gamma_k^\text{MF} \right)^{-2}}\left( 1+\Gamma_k^\text{MF} \right)^2 \right].
	\end{align}
	Since the order of magnitude for the bandwidth\footnote{In the operational wireless systems, the MIMO-related techniques are generally used at the narrowband (subcarriers), hence the order of magnitude on basic scheduling resource is of kHz such as the resource block (180 kHz) in the LTE-related systems.} is generally of kHz, we can obtain
	\begin{align}
	& -Q^{-1}(\epsilon_k)-\eta_j \sqrt{B^*}\sqrt{1-\left( 1+\Gamma_k^\text{MF} \right)^{-2}}\left( 1+\Gamma_k^\text{MF} \right)^2 \notag \\
	&\qquad \geqslant -Q^{-1}(\epsilon_k)-\eta_j \sqrt{B^*}\dfrac{\Gamma_k^\text{MF}}{1+\Gamma_k^\text{MF}} = \dfrac{-Q^{-1}(\epsilon_k)\left( 1+\Gamma_k^\text{MF} \right)-\eta_j \sqrt{B^*}\Gamma_k^\text{MF}}{1+\Gamma_k^\text{MF}}>0,
	\end{align}
\else
	\typeout{The twocolumn mode.}
	\begin{figure*}[!t]
		\centering
		\begin{align}
		\label{diffh}
		\dfrac{\partial h_k^\text{MF}(\myvec{p})}{\partial p_k} = \dfrac{\sqrt{B^*}\left( \dfrac{\Gamma_k^\text{MF}}{p_k}+\dfrac{(\Gamma_k^\text{MF})^2}{Mp_k} \right)}{\sqrt{1-\left( 1+\Gamma_k^\text{MF} \right)^{-2}}\left( 1+\Gamma_k^\text{MF} \right)^3 \ln 2} \left[ -Q^{-1}(\epsilon_k)-\eta_j \sqrt{B^*}\sqrt{1-\left( 1+\Gamma_k^\text{MF} \right)^{-2}}\left( 1+\Gamma_k^\text{MF} \right)^2 \right].
		\end{align}
		\hrulefill
		\vspace*{-10pt}
	\end{figure*}
	To solve Sub-problem~\ref{subpro1}, the monotonicity of $ h_k(\myvec{p}) = f_k(\myvec{p})-\eta_j g_k(\myvec{p}) $ is first studied. The proof of the ZF is similar to that of the MF, hence only the case of the MF is shown here. The derivative $ \partial h_k^\text{MF}(\myvec{p})/\partial p_k =\partial f_k^\text{MF}(\myvec{p})/\partial p_k - \eta_j \partial g_k^\text{MF}(\myvec{p})/\partial p_k $ is given by \eqref{diffh}. Since the order of magnitude for the bandwidth\footnote{In the operational wireless systems, the MIMO-related techniques are generally used at the narrowband (subcarriers), hence the order of magnitude on basic scheduling resource is of kHz such as the resource block (180 kHz) in the LTE-related systems.} is generally of kHz, we can obtain
	\begin{align}
	& -Q^{-1}(\epsilon_k)-\eta_j \sqrt{B^*}\sqrt{1-\left( 1+\Gamma_k^\text{MF} \right)^{-2}}\left( 1+\Gamma_k^\text{MF} \right)^2 \notag \\
	&\qquad \geqslant -Q^{-1}(\epsilon_k)-\eta_j \sqrt{B^*}\dfrac{\Gamma_k^\text{MF}}{1+\Gamma_k^\text{MF}} \notag \\
	&\qquad = \dfrac{-Q^{-1}(\epsilon_k)\left( 1+\Gamma_k^\text{MF} \right)-\eta_j \sqrt{B^*}\Gamma_k^\text{MF}}{1+\Gamma_k^\text{MF}}>0,
	\end{align}
\fi
which implies that the function $ h_k^\text{MF}(\myvec{p})=f_k^\text{MF}(\myvec{p})-\eta_j g_k^\text{MF}(\myvec{p}) $ is increasing. The theorem capable of achieving the optimal max-min power allocation is presented as follows~\cite{Zhao2017}.

\begin{theorem}
\label{theo5}
In order to maximize the minimum objective function value of \eqref{subproblem1}, all VUEs should have the same objective value, hence we have
\begin{align}
\label{theorem5}
h_1(\myvec{p}) = h_2(\myvec{p}) = \cdots = h_K(\myvec{p}) \triangleq H_{\myvec{p}}.
\end{align}
\end{theorem}

\begin{IEEEproof}
\eqref{theorem5} can be obtained by the method of contradiction. Let $ \myvec{p}^*=[p_1^*, \cdots, p_K^*]^\text{T} $ be the optimal power allocation of Sub-problem~\ref{subpro1}, which does not satisfy~\eqref{theorem5}. Without loss of generality, let $ h_1(p_1^*) = \max_{k}\left\lbrace h_k(p_k^*) \right\rbrace $ and $ h_K(p_K^*) = \min_{k}\left\lbrace h_k(p_k^*) \right\rbrace $. As $ h_k(\myvec{p}) $ is monotonically increasing, we should have $ \mu \in (0,p_1^*) $ satisfying
\begin{align}
h_1(p_1^*) > h_1(p_1^*-\mu) = h_K(p_K^*+\mu) > h_K(p_K^*).
\end{align}
Hence, a new power allocation $ \myvec{p}^*_\text{N}=[p_1^*-\mu, \cdots, p_K^*+\mu]^\text{T} $ satisfying \eqref{subproblem1con} is constructed. In other words, the minimum objective function value has been increased, which contradicts to the max-min criterion. Hence the theorem is proved.
\end{IEEEproof}

Based on Theorem~\ref{theo5}, an iterative procedure described by Algorithm~\ref{alg2} is proposed for solving Sub-problem~\ref{subpro1}. According to $ \lim_{i \rightarrow \infty}( | h_{k_{i}^\text{m}}^{i}(\myvec{p}_{i})-h_{k_{i}^\text{M}}^{i}(\myvec{p}_{i}) | )=0 $, we can intuitively find that Algorithm~\ref{alg2} is capable of converging to the same objective function value $ H_{\myvec{p}} $. Meanwhile, the optimality can be guaranteed by Theorem~\ref{theo5}. In the initialization phase of Algorithm~\ref{alg2}, the objective values of all VUEs are calculated for the equal power allocation. As the iterations proceed, Algorithm~\ref{alg2} will accordingly increase the minimum power and reduce the maximum power every time until we have $ \max_{k}\left\lbrace \myvec{H}_i \right\rbrace - \min_{k}\left\lbrace \myvec{H}_i \right\rbrace \leqslant \zeta_\text{S} $. To avoid the ping-pong phenomenon between $ k_{i}^\text{m} $ and $ k_{i}^\text{M} $, four judgement conditions are listed in Algorithm~\ref{alg2} to adjust the allocation step length $ \mu_i $.

\subsubsection{Algorithm Complexity}
First of all, the optimal solution of Problem~\ref{pro1} is directly achieved by the quadratic formula. Furthermore, since this optimal solution is independent of the number of VUEs and it is a closed-form solution, the complexity of solving Problem~\ref{pro1} can be ignored. With respect to solving Problem~\ref{pro2}, the convergence and optimality are guaranteed by Theorem~\ref{theo4} and \ref{theo5}, while the complexity is determined by Stage 2 of Algorithm~\ref{alg1}, and Algorithm~\ref{alg2}. Based on the maximum iterative tolerance $ \zeta_\text{P} $ of Algorithm~\ref{alg1}, the associated precision is given by $ Z = \log_{10}(\zeta_\text{P}^{-1}) $-digit. Thus, the complexity order of Algorithm~\ref{alg1} is $ \mathcal{O}(\log_2 Z) $~\cite{Borweinbook,Zappone2015}. On the other hand, because $ | h_k(p_k)-H_{\myvec{p}} | \leqslant \frac{\zeta_\text{S}}{2} $ is a sufficient condition for satisfying the termination condition $  \max_{k}\left\lbrace \myvec{H}_i \right\rbrace - \min_{k}\left\lbrace \myvec{H}_i \right\rbrace \leqslant \zeta_\text{S} $, the complexity of Algorithm~\ref{alg2} is on the order of $ \mathcal{O}( \sum_{k=1}^{K} \log_2[ h_k^{-1}( H_{\myvec{p}}+\frac{\zeta_\text{S}}{2} ) - h_k^{-1}( H_{\myvec{p}}-\frac{\zeta_\text{S}}{2} ) ]^{-1} ) $~\cite{Zhao2019,Zhao2017}. Hence, the total complexity order is $ \mathcal{O}( \log_2 Z \sum_{k=1}^{K} \log_2[ h_k^{-1}( H_{\myvec{p}}+\frac{\zeta_\text{S}}{2} ) - h_k^{-1}( H_{\myvec{p}}-\frac{\zeta_\text{S}}{2} ) ]^{-1} ) $~\cite{Cheung2013}.

Finally, due to the handover in high mobility VNETs, we discuss the scalability of the proposed algorithms.
\begin{remark}[Extension of Algorithms]
The proposed algorithms depend on the road-traffic density and the location information of all VUEs. In general, the traffic control center can use roadside monitoring sensors or real-world traffic datasets to predict and periodically report the vehicular density and location information~\cite{Lv2015,Polson2017}. For multi-cell scenarios, both of the density and location information can be shared via wired links between the traffic control center and all the base stations. Therefore, as long as the density and location information are available, our proposed algorithms can be readily extended to the multi-cell scenarios. In addition, due to the wired links, the sharing of the density and location information do not bring the extra signal overhead for radio access networks.
\end{remark}

\section{Simulation Results}
\label{sec:Simulation}

\subsection{Simulation Setup}

\begin{table}[!t]
	\setlength{\extrarowheight}{1pt}
	\centering
	\caption{Basic Simulation Parameters}
	\begin{tabular}{ l | r }
		\hline
		\textbf{Parameter} & \textbf{Value} \\
		\hline
		\hline
		BS distance $ d_\text{B} $ & 20~\myunit{m} \\
		\hline
		Road length $ d_\text{R} $ & 200~\myunit{m} \\
		\hline
		Maximum road-traffic density $ \rho_\text{m} $ & 0.15 \\
		\hline
		Free flow speed $ v_\text{F} $ & 80~\myunit{km/h} \\
		\hline
		Estimation accuracy $ \chi_k, \forall k $ & 0.8 \\
		\hline
		Carrier frequency $ f_\text{C} $ & 2~\myunit{GHz} \\
		\hline
		Constant $ \theta $ & $ 10^{-3} $ \\ 
		\hline
		Path loss exponent $ \alpha $ & 3.8 \\
		\hline
		Noise power spectral density $ N_0 $ & -130~\myunit{dBm/Hz} \\
		\hline
		Transmitted signal power spectral density $ P_0 $ & -10~\myunit{dBm/Hz} \\
		\hline
	\end{tabular}
	\label{table_1}
\end{table}

In the simulations, the large-scale fading of the $ k $-th VUE is given by $ \beta_k = \theta [ ( d_k-d_\text{R}/2 )^2+d_\text{B}^2 ]^{-\alpha/2} $, where $ \theta $ is a constant related to the antenna gain and carrier frequency, $ \alpha $ is the path loss exponent, and $ d_k $ is the distance between the $ k $-th VUE and the starting point of the road. The noise power is calculated by $ \sigma^2= N_0B $, where $ N_0 $ and $ B $ are the noise power spectral density and system bandwidth, respectively. Furthermore, all VUEs are uniformly distributed on the one-way road, namely $ d_k \sim \text{unif}(0,d_\text{R}) $. Since the average length of a vehicle is 6.5~\myunit{m} (already plus 50\% of additional safety distance), the maximum road-traffic density is set to $ \rho_\text{m} = 0.15 $. In the following simulation results, PCSI and IPCSI represent the perfect and imperfect channel state information, respectively. Finally, all other basic simulation parameters are listed in Table~\ref{table_1}.

\subsection{Simulation Results for the Maximum Achievable Rate and Transmission Latency}

\subsubsection{Tradeoff}

\begin{figure}[!t]
	\centering
	\includegraphics[scale=0.45]{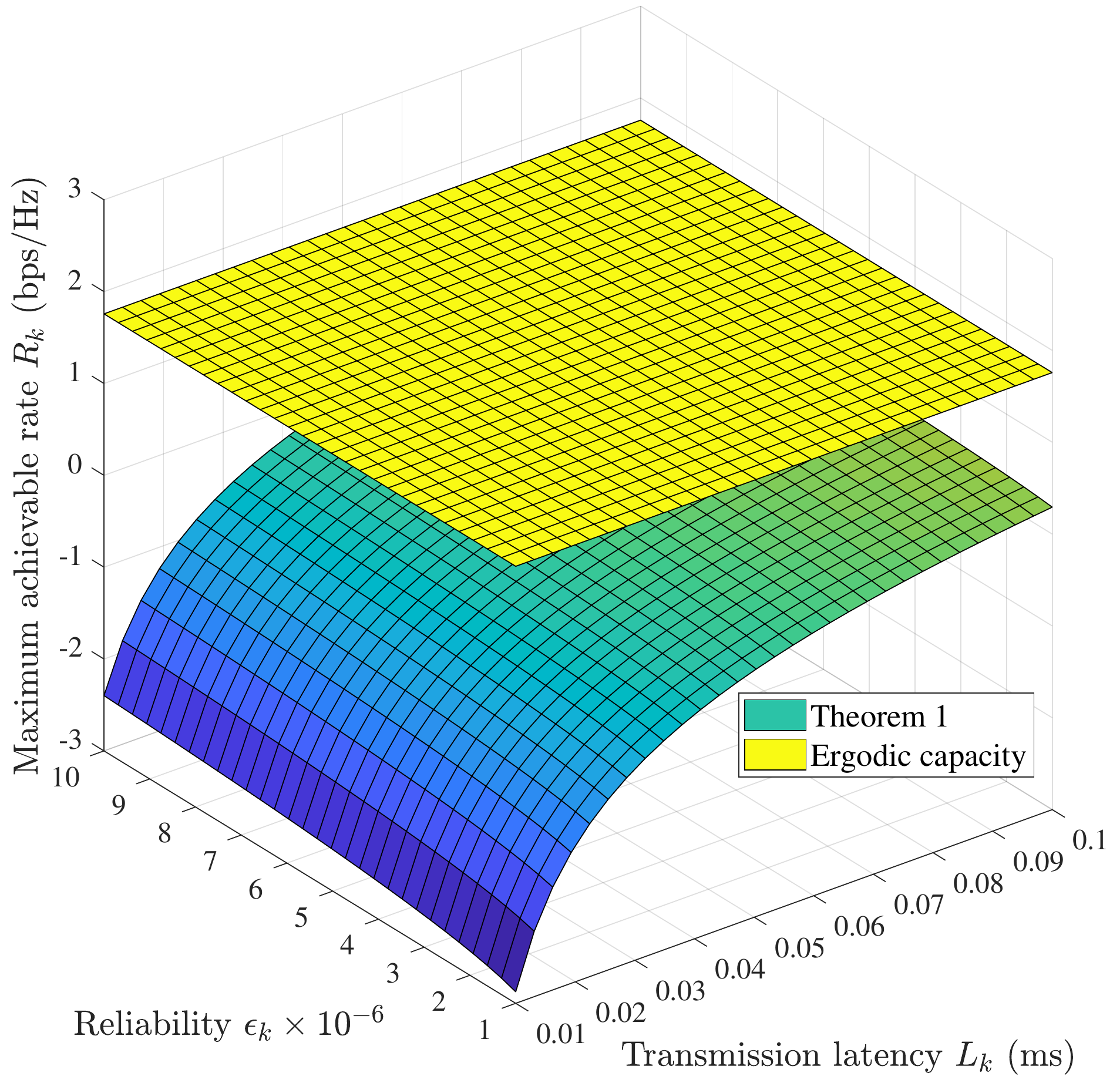}
	\caption{Tradeoff among maximum achievable rate, transmission latency and reliability in \eqref{Shannon} and \eqref{theorem1} based on the MF precoder, with $ \rho = 0.05 $, $ M =300 $, $ B = 200 $~\myunit{kHz} and $ P_\text{B}=10 $~\myunit{dBW}.}
	\label{fig_RLR_MF}
\end{figure}

Upon using the equal power allocation (EPA) and MF precoder, Fig.~\ref{fig_RLR_MF} illustrates the tradeoff of a VUE among the maximum achievable rate, transmission latency and reliability for the case of IPCSI. As shown in Fig.~\ref{fig_RLR_MF}, regardless of what the latency and reliability are, the ergodic capacity remains a constant. By contrast, Theorem~\ref{theo1} indicates that some rate regions are not achievable (negative) because of the ultra-high reliability and ultra-low latency. Moreover, the maximum achievable rate given by Theorem~\ref{theo1} is lower than the ergodic capacity, which confirms that V2I URLLCs are indeed possible, but only at the cost of a reduced rate. Finally, Fig.~\ref{fig_RLR_MF} shows that the transmission latency can only be optimized within a reasonable range. The results of the ZF are similar to those of the MF, thus they are omitted here due to space limitations.

\subsubsection{Tightness}

\begin{figure}[!t]
	\centering
	\includegraphics[scale=0.3]{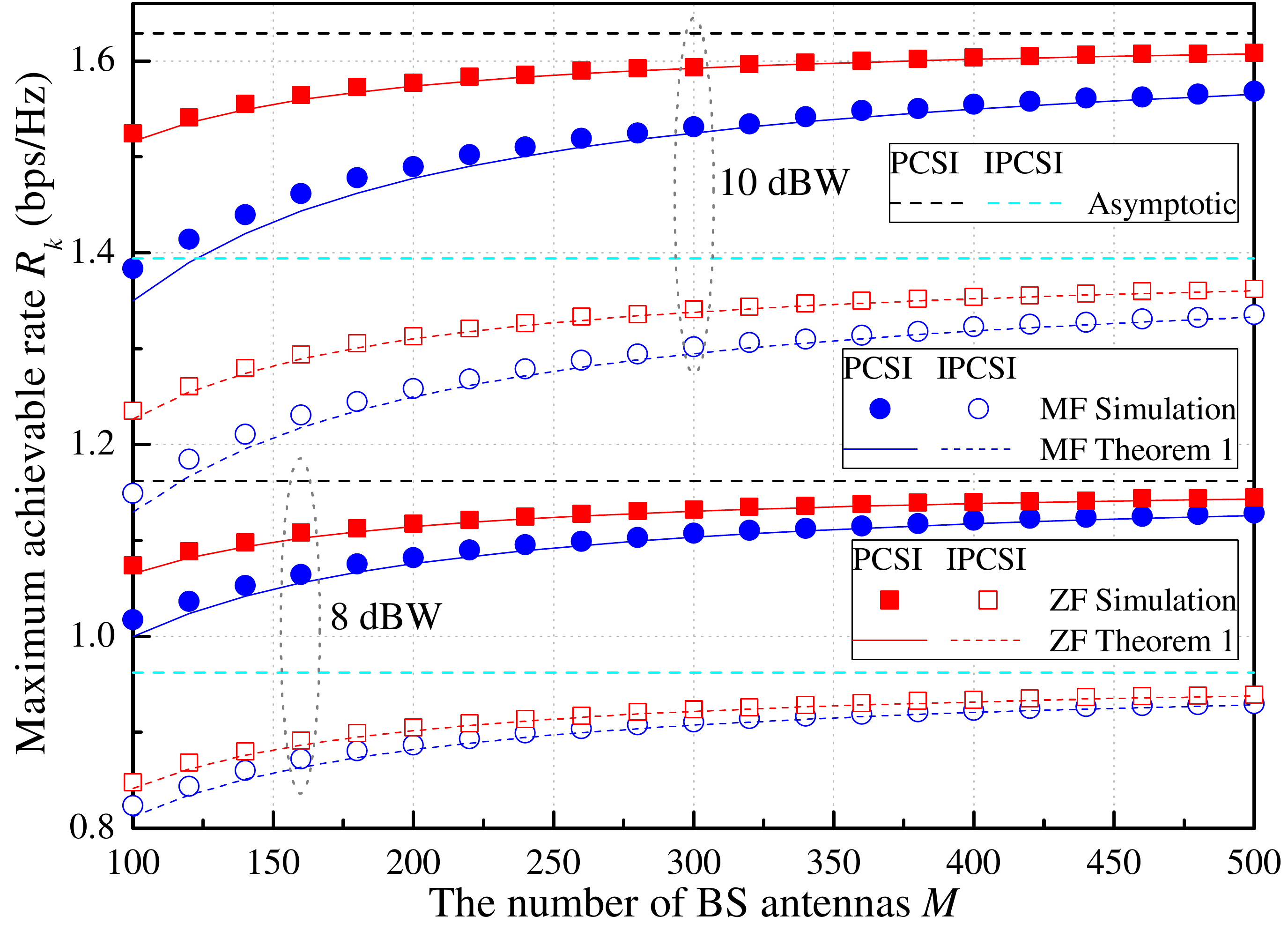}
	\caption{Maximum achievable rate versus the number of antennas in \eqref{theorem1}, with $ \rho = 0.05 $, $ B = 200 $~\myunit{kHz} and $ L_k = 1 $~\myunit{ms}.}
	\label{fig_MvsR}
\end{figure}

\begin{figure}[!t]
	\centering
	\includegraphics[scale=0.3]{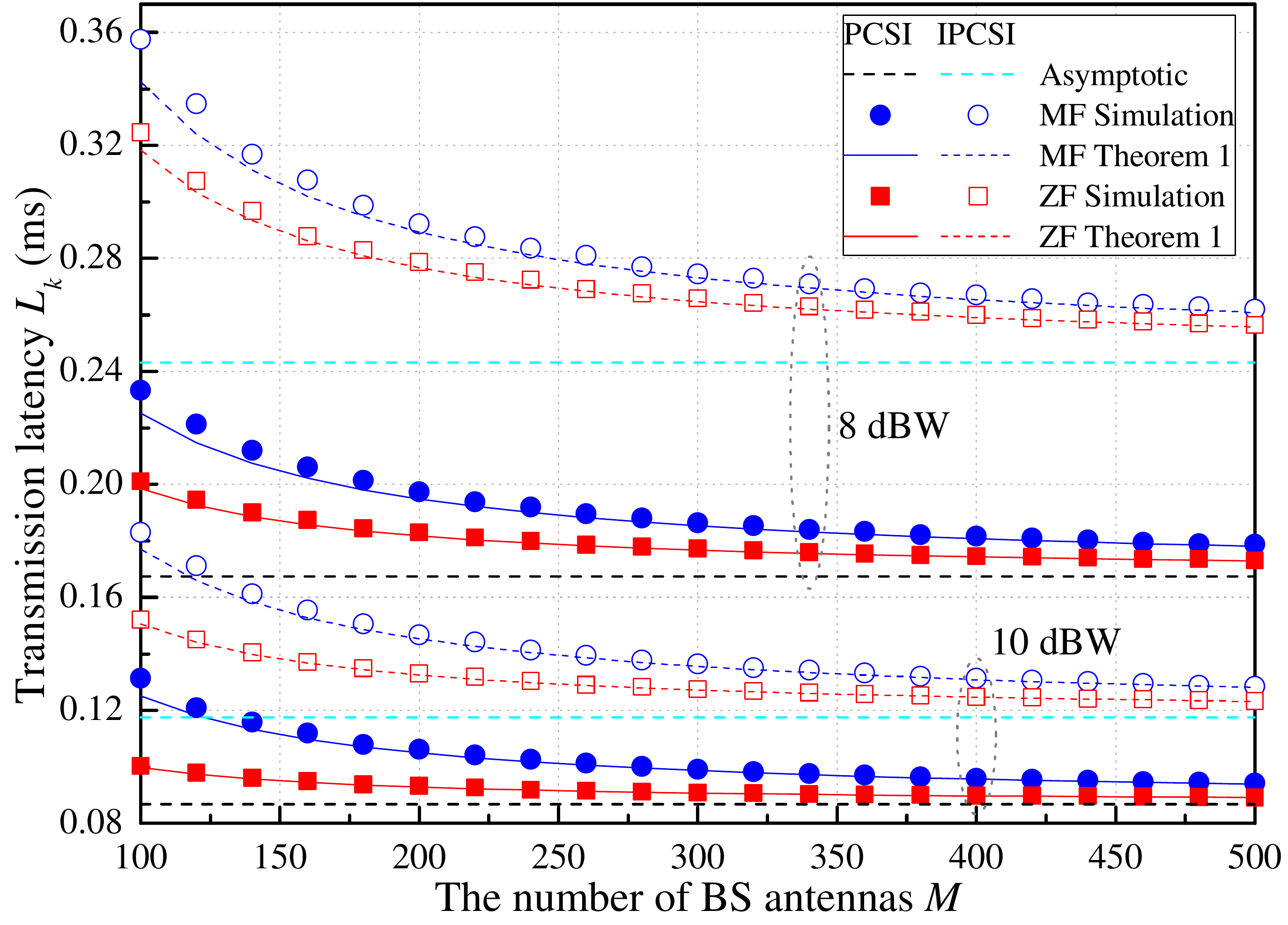}
	\caption{Transmission latency versus the number of antennas in \eqref{metric},  with $ \rho = 0.05 $, $ B = 200 $~\myunit{kHz} and $ R_k = 100 $~\myunit{kbps}.}
	\label{fig_MvsL}
\end{figure}

Upon adopting the EPA schemes, Fig.~\ref{fig_MvsR} and \ref{fig_MvsL} illustrate the maximum achievable rate and transmission latency versus the number of antennas, respectively. The reliability threshold $ \epsilon_k $ of these two figures is set to $ 10^{-6} $. The asymptotic results ($ M \rightarrow \infty $) are calculated from \eqref{theorem1} and \eqref{metric} with $ \Gamma^\infty_k=p_k \chi_k \beta_k / \sigma^2 $ ($ \chi_k = 1 $ for PCSI). For the PCSI and IPCSI, Fig.~\ref{fig_MvsR} and \ref{fig_MvsL} indicate that upon increasing the number of antennas, the maximum achievable rate increases, while the latency of a VUE decreases. The approximate results of Theorem~\ref{theo1} and Corollary~\ref{coro1} are close to the simulation results, and both of them tend to the asymptotic results as the number of antennas increases. Because of the channel estimation error, the case of IPCSI is always worse than the case of PCSI. Furthermore, both the rate and latency of the ZF are better than those of the MF in the case of PCSI or IPCSI, due to the higher SINR $ \Gamma_k $. Finally, these two figures suggest that increasing the total transmitted power improve both the rate and latency. Based on the tightness of Corollary~\ref{coro1}, the simulation results of Problem~\ref{pro1} and~\ref{pro2} will be discussed in the following subsections.

\subsection{Simulation Results for Problem~\ref{pro1}}

\begin{figure}[!t]
	\centering
	\includegraphics[scale=0.3]{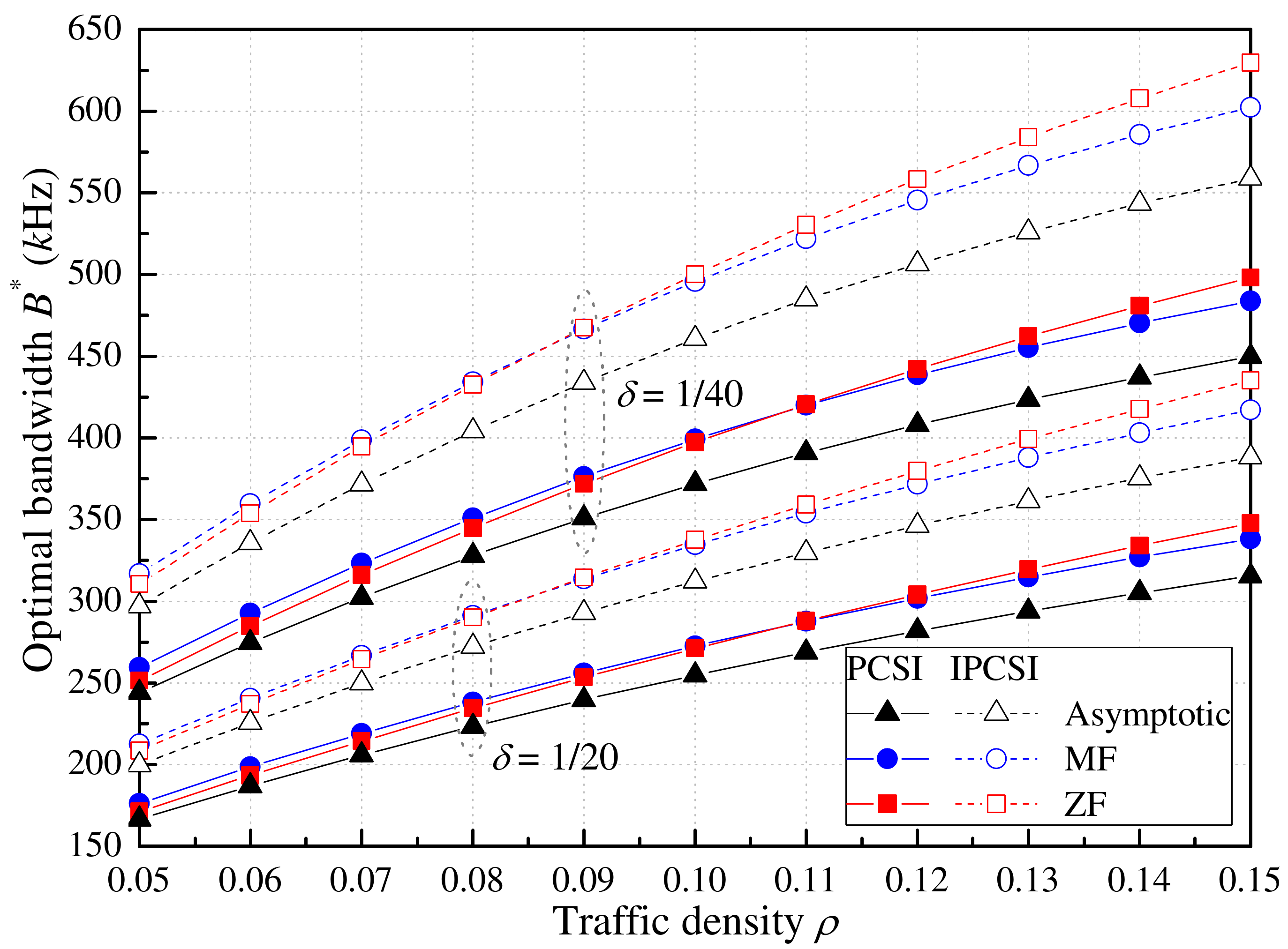}
	\caption{Optimal system bandwidth versus road-traffic density in \eqref{optimalB}, with $ M=300 $, $ R_\text{W} = 100 $~\myunit{kbps} and $ \epsilon_\text{W} = 10^{-6} $.}
	\label{fig_rhovsB}
\end{figure}

\begin{figure}[!t]
	\centering
	\includegraphics[scale=0.3]{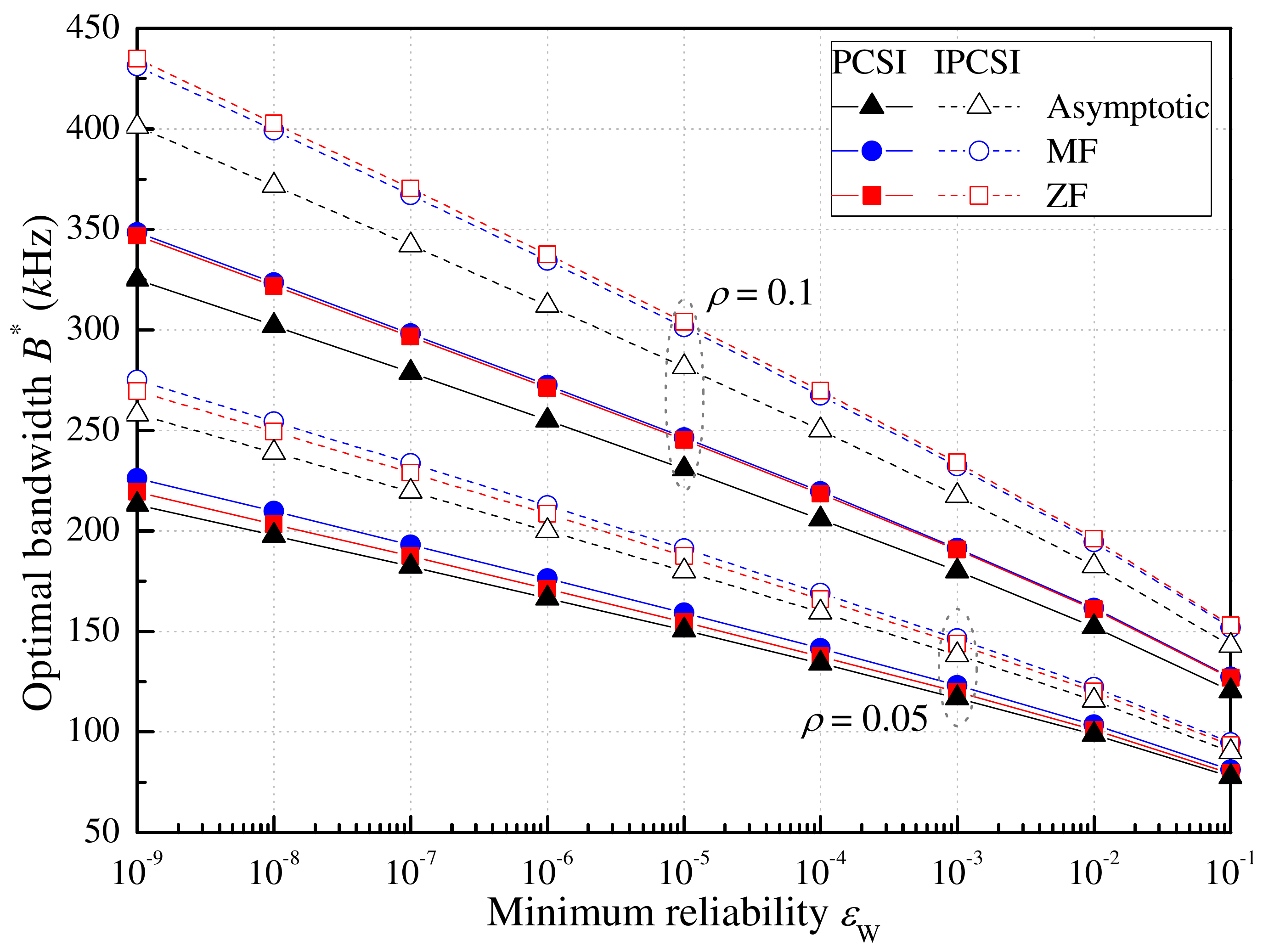}
	\caption{Optimal system bandwidth versus minimum reliability in \eqref{optimalB}, with $ M=300 $, $ R_\text{W} = 100 $~\myunit{kbps} and $ \delta = 1/20 $.}
	\label{fig_epsilonvsB}
\end{figure}

Based on the PCSI and IPCSI, Fig.~\ref{fig_rhovsB} illustrates the optimal system bandwidth versus the road-traffic density, where the MF, ZF and asymptotic results are calculated by substituting \eqref{worstGam} and $ \Gamma^\infty_\text{W} = P_0 \chi_\text{th} \beta_\text{W} / (N_0 \rho d_\text{R}) $ into \eqref{optimalB}. In order to keep the worst-case latency below a certain threshold, Fig.~\ref{fig_rhovsB} shows that with the road-traffic density increasing, the optimal system bandwidth also increases. This means that the more VUEs are supported, the more bandwidth is required. In the region of low road-traffic density, the optimal bandwidth of the ZF is lower than that of the MF, whereas in the region of high road-traffic density, the required bandwidth of the ZF is higher than that of the MF due to the lower SINR. Moreover, the increase of the threshold $ \delta $ requires more bandwidth. Compared to the case of PCSI, the case of IPCSI requires more bandwidth to satisfy the worst-case latency requirement. Finally, when the number of antennas tends to infinity, the bandwidth is only reduced by at most 75~\myunit{kHz} (IPCSI at $ \delta = 1/40 $ with $ \rho = 0.15 $), which implies that the V2I URLLC system does not require an excessive numbers of antennas.

Fig.~\ref{fig_epsilonvsB} illustrates the optimal system bandwidth versus the reliability for the two road-traffic densities considered. As shown in Fig.~\ref{fig_epsilonvsB}, upon relaxing the minimum reliability requirement, the optimal system bandwidth decreases, regardless of the CSI. The gap between the MF results and ZF results is in line with the trend seen in Fig.~\ref{fig_rhovsB} under these two types of road-traffic densities. Additionally, by comparing the results of $ \epsilon_\text{W} = 10^{-6} $ to those of $ 10^{-9} $, the optimal bandwidth is only increased by at most 75~\myunit{kHz} (IPCSI at $ \rho = 0.1 $), which suggests that the reliability can be significantly improved by only modestly increasing the bandwidth.

\subsection{Simulation Results for Problem~\ref{pro2}}

\subsubsection{Convergence}

\begin{figure}[!t]
	\centering
	\subfloat[Convergence of Algorithm~\ref{alg1} in \eqref{optimaleta} and Theorem~\ref{theo4}.]{\includegraphics[scale=0.3]{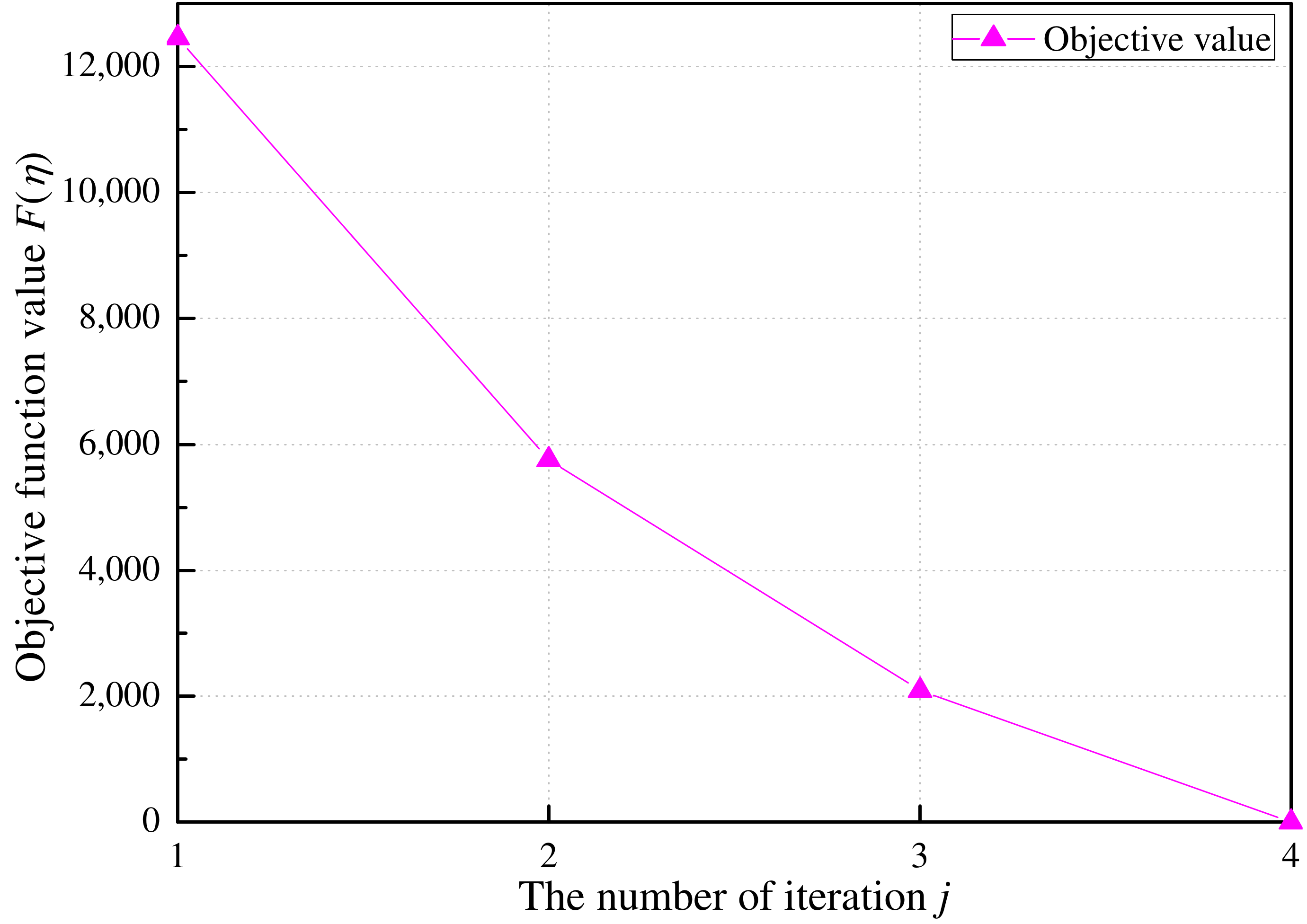}\label{fig_Cov_Alg1}}
	\hfil
	\subfloat[Convergence of Algorithm~\ref{alg2} in Theorem~\ref{theo5}.]{\includegraphics[scale=0.3]{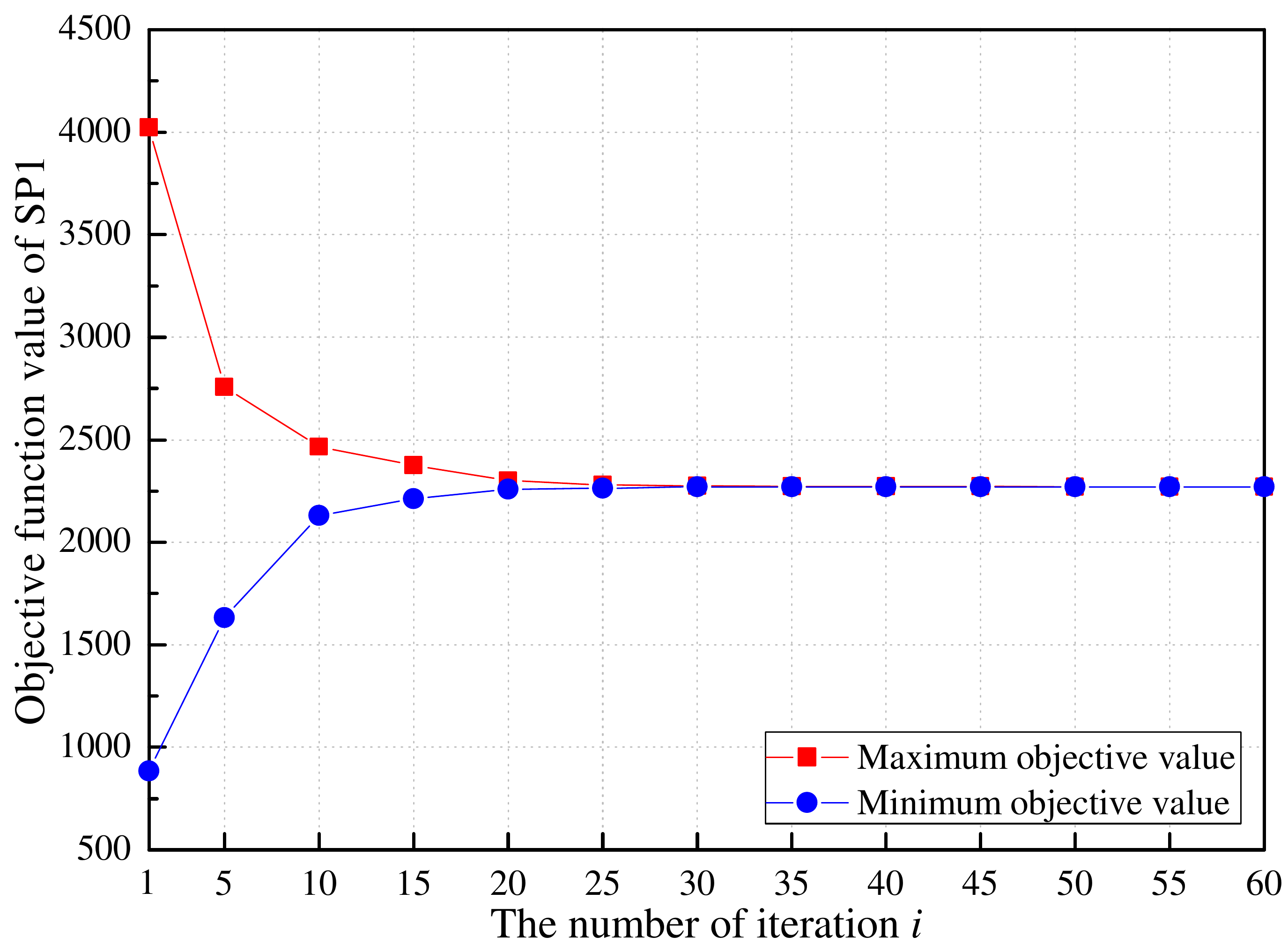}\label{fig_Cov_Alg2}}
	\caption{Convergence of the two proposed algorithms with $ \zeta_\text{P} = \zeta_\text{S} = 10^{-2} $, $ \eta_0 = -3 \times 10^{-2} $ and $ \mu_0 = P_\text{B}/(2 \rho d_\text{R}) $.}
	\label{fig_Cov}
\end{figure}

Based on a specific channel realization in the simulations, Fig.~\subref*{fig_Cov_Alg1} and \subref*{fig_Cov_Alg2} illustrate the convergence of the two proposed algorithms, respectively. Fig.~\subref*{fig_Cov_Alg1} indicates that as the number of iterations increases, the auxiliary function $ F(\eta) $ converges to zero in Algorithm~\ref{alg1}, as stated by Theorem~\ref{theo4}. On the other hand, as shown in Fig.~\subref*{fig_Cov_Alg2}, as the number of iterations increases, the maximum objective function value decreases and the minimum objective function value increases. All objective function values finally converge to a constant, which characterizes the convergence of Algorithm~\ref{alg2}.

\subsubsection{Transmission Latency}

\begin{figure}[!t]
	\centering
	\includegraphics[scale=0.3]{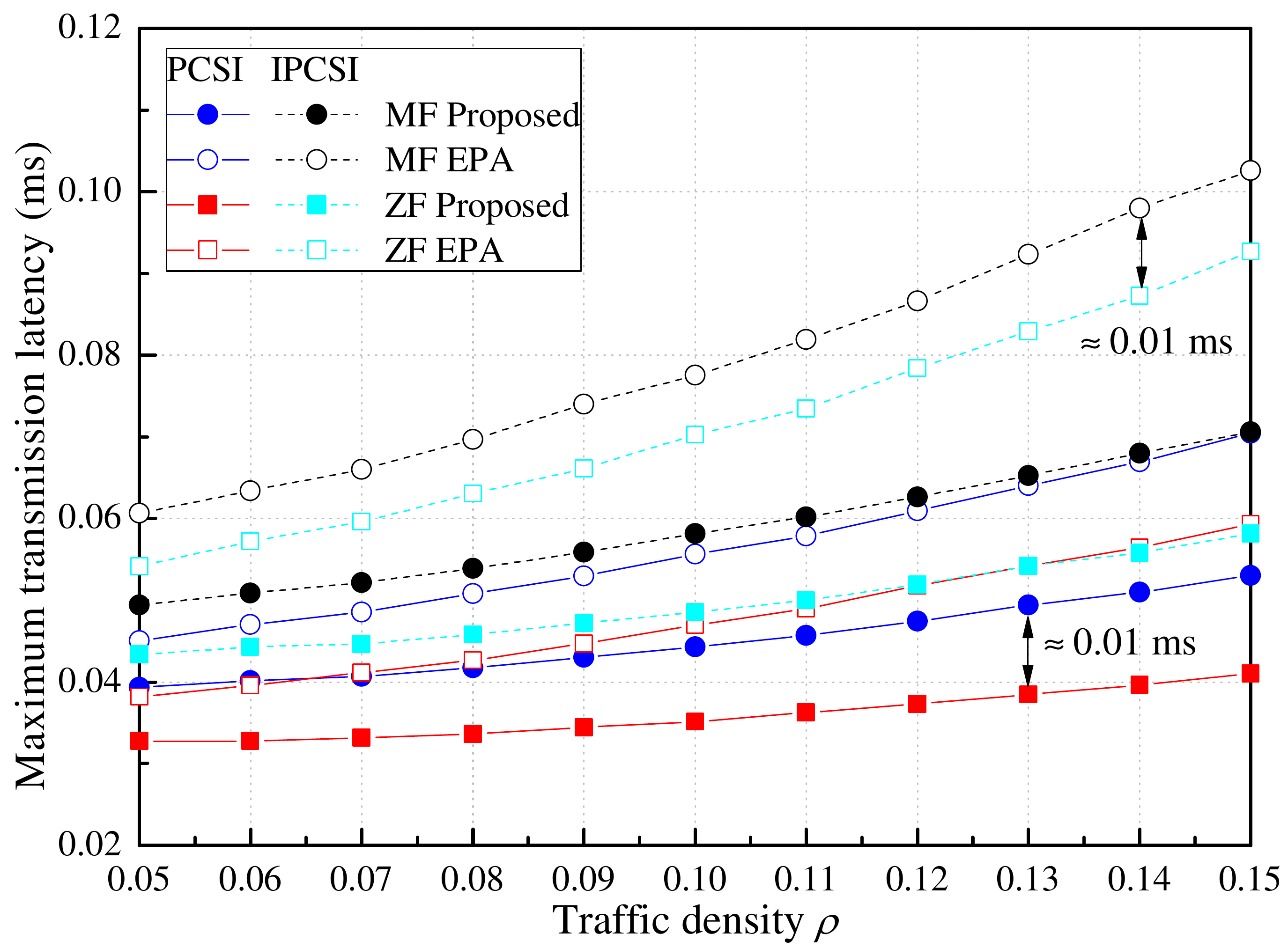}
	\caption{Maximum transmission latency versus road-traffic density solved by Algorithm~\ref{alg1} and \ref{alg2}, with $ M =300 $, $ \delta = 1/20 $, $ R_k = 100 $~\myunit{kbps} and $ \epsilon_k = 10^{-6} $ for all VUEs.}
	\label{fig_rhovsL}
\end{figure}

\begin{figure}[!t]
	\centering
	\includegraphics[scale=0.3]{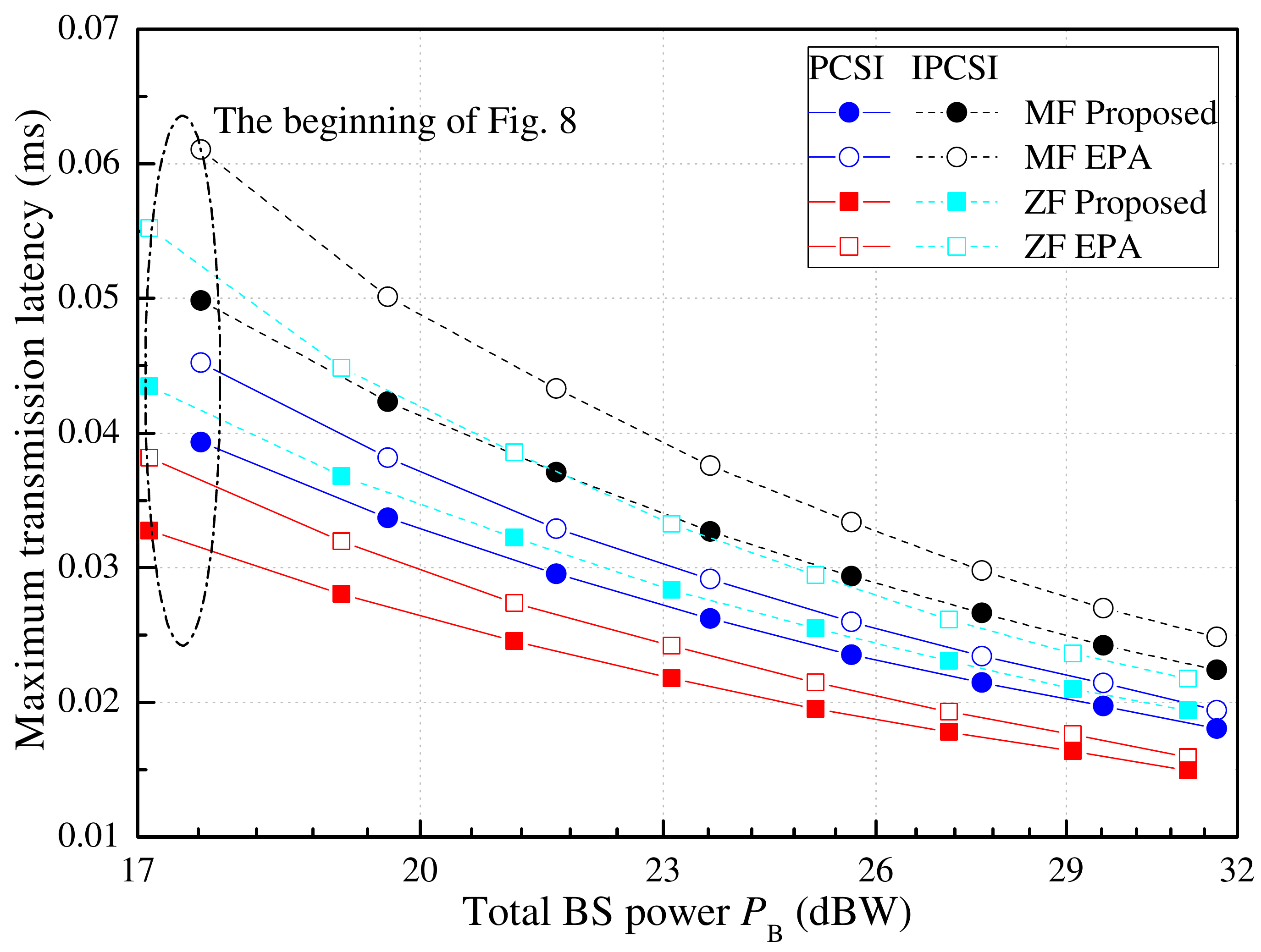}
	\caption{Maximum transmission latency versus total transmitted power solved by Algorithm~\ref{alg1} and \ref{alg2}, with $ M =300 $, $ \delta = 1/20 $, $ \rho = 0.05 $, $ R_k = 100 $~\myunit{kbps} and $ \epsilon_k = 10^{-6} $ for all VUEs.}
	\label{fig_PBvsL}
\end{figure}

Fig.~\ref{fig_rhovsL} illustrates the maximum transmission latency among all VUEs versus the road-traffic density. Observe that when the road-traffic density increases, the maximum transmission latency of both the proposed scheme and of the EPA scheme increase accordingly. Upon comparing the MF precoder to the ZF precoder, regardless of having either PCSI or IPCSI, the performance of the ZF is better than that of the MF both for the proposed scheme and for the EPA scheme. Fig.~\ref{fig_rhovsL} also indicates that the proposed scheme performs better than the EPA scheme both for the MF precoder and for the ZF precoder in the case of either PCSI or IPCSI. Furthermore, the performance gap between the MF and ZF is always about 0.01~\myunit{ms}. Naturally, the channel estimation error increases latency. Nevertheless, the IPCSI results of the proposed scheme are close to the PCSI results of the EPA scheme both in the case of the MF and the ZF precoder. Finally, the slope of the proposed scheme is lower for both the PCSI and IPCSI, which implies that the proposed scheme is not sensitive to the road-traffic density. Hence, the proposed algorithms can get the effect of semi-persistent scheduling for the V2I URLLC system.

Given $ \rho = 0.05 $ and increase the total power based on $ P_0B^* $, Fig.~\ref{fig_PBvsL} quantifies the maximum transmission latency among all VUEs versus the total transmitted power. As shown in Fig.~\ref{fig_PBvsL}, the maximum transmission latency decreases for both the proposed scheme and for the EPA scheme, as the total power increases. The performance of the ZF is better than that of the MF both for the proposed scheme and for the EPA scheme in the case of either PCSI or IPCSI. Additionally, the gap between the ZF relying on IPCSI and the MF using PCSI is relatively small for the proposed scheme. Hence, the ZF precoder is the better choice for V2I URLLCs, as evidenced both by Fig.~\ref{fig_rhovsL} and Fig.~\ref{fig_PBvsL}.

\section{Conclusions}
\label{sec:Conclusion}

In order to reduce the signaling overhead, this paper studied the problem of minimizing the transmission latency in the downlink of the V2I URLLC system based on the road-traffic density and large-scale fading CSI. The expression of transmission latency was first derived for the MF precoder and ZF precoder based on both perfect and imperfect CSI. Then, a two-stage radio resource allocation problem was formulated relying on our twin-timescale perspective. More particularly, the first stage optimized the worst-case transmission latency by adjusting the system's bandwidth based on a long-term timescale, while the second stage minimized the maximum transmission latency among all VUEs by judiciously allocating the total power on a short-term timescale basis. For optimally solving our two-stage problem formulated, a moderate-complexity twin-timescale radio resource allocation algorithm was conceived. Our simulation results showed that the proposed resource allocation scheme significantly reduced the maximum transmission latency and that the ZF precoder was the better choice for the V2I URLLC system considered.

\appendices
\section{Proof of Theorem~\ref{theo1}}
\label{app:theo1}

Recall from \eqref{approx} that the ergodic rate of the $ k $-th VUE is given by
\begin{align}
& R_k(\gamma_k,L_k,\epsilon_k) \approx \mathbb{E}\left\lbrace B\left[ \log_2(1+\gamma_k)-\sqrt{\dfrac{V_k}{L_kB}}Q^{-1}(\epsilon_k) \right] \right\rbrace \notag \\
&\quad= \mathbb{E}\left[ B \log_2\left( 1+\dfrac{1}{\gamma_k^{-1}} \right) \right] - \mathbb{E}\left[ \sqrt{\dfrac{BV_k}{L_k}}Q^{-1}(\epsilon_k) \right].
\end{align}
Let $ u_k=\log_2(1+1/\gamma_k^{-1}) $ denote the spectral efficiency of the $ k $-th VUE, then $ V_k $ can be rewritten as
\begin{align}
V_k = \left( 1- \dfrac{1}{(1+\gamma_k)^2} \right) \left( \log_2 \myexp \right)^2 = \left( 1-2^{-2u_k} \right) \left( \log_2 \myexp \right)^2,
\end{align}
and we have
\begin{align}
\label{Rtemp}
& R_k(\gamma_k,L_k,\epsilon_k) \approx B\mathbb{E}\left[ \log_2\left( 1+\dfrac{1}{\gamma_k^{-1}} \right) \right] \notag \\
&\qquad \qquad - \log_2 \myexp \sqrt{\dfrac{B}{L_k}}Q^{-1}(\epsilon_k)\mathbb{E}\left[ \sqrt{1-2^{-2u_k}} \right].
\end{align}

Since the function $ a(x)=\log_2(1+1/x), \forall x>0 $ is convex, Jensen's inequality gives
\begin{align}
\label{lowerb}
U_k\triangleq\mathbb{E}\left[ u_k \right] &\geqslant U_k^\text{L}\triangleq\log_2\left( 1+\dfrac{1}{\mathbb{E}[\gamma_k^{-1}]} \right).
\end{align}
Substituting $ \mathbb{E}\left[ \Omega_\text{B}^\text{MF} \right] = \frac{1}{M} $, $ \mathbb{E}\left[ \Omega_\text{G}^\text{MF} \right] = \frac{1}{M-1} $ and $ \mathbb{E}\left[ \Omega_\text{G}^\text{ZF} \right] = \frac{1}{M-K} $ into \eqref{SINR}, we obtain
\begin{align}
U_k^\text{L} =
\begin{cases}
\log_2\left( 1+\dfrac{Mp_k}{P_\text{B}-p_k+\varphi_k} \right), &\text{for MF},\\
\log_2\left( 1+p_k\varphi_k \right), &\text{for ZF},
\end{cases}
\end{align}
where
\begin{align}
\varphi_k =
\begin{cases}
\dfrac{P_\text{B}\beta_k\left( 1-\chi_k \right)+M\sigma^2}{\chi_k\beta_k}\dfrac{M}{M-1}, &\text{for MF},\\
\dfrac{\chi_k\beta_k\left( M-K \right)}{P_\text{B}\beta_k\left( 1-\chi_k \right)+M\sigma^2}, &\text{for ZF}.
\end{cases}
\end{align}
Note that for $ M\rightarrow\infty $, we have $ U_k^\text{L}=\log_2(1+\frac{p_k\chi_k\beta_k}{\sigma^2}) $. On the other hand, due to the channel hardening phenomenon, we arrive at $ \lim_{M\rightarrow\infty}U_k=\log_2(1+\dfrac{p_k\chi_k\beta_k}{\sigma^2}) $, which indicates that $ U_k^\text{L} $ is tight for $ U_k $ when $ M $ is large. Therefore, $ U_k^\text{L} $ can be used for approximating the first term of \eqref{Rtemp}, i.e., 
\begin{align}
\label{firstterm}
\mathbb{E}\left[ \log_2\left( 1+\dfrac{1}{\gamma_k^{-1}} \right) \right] \approx U_k^\text{L}, M \gg 1.
\end{align}

Similarly, since the function $ b(x)=\sqrt{1-2^{-2x}}, \forall x>0 $ is concave, with Jensen's inequality,
\begin{align}
\mathbb{E}\left[ \sqrt{1-2^{-2u_k}} \right] &\leqslant \sqrt{1-2^{-2\mathbb{E}[u_k]}}.
\end{align}
Based on the channel hardening phenomenon, we arrive at
\begin{align}
\lim_{M\rightarrow\infty}\mathbb{E}\left[ \sqrt{1-2^{-2u_k}} \right]&=\sqrt{1-\left( 1+\dfrac{p_k\chi_k\beta_k}{\sigma^2} \right)^{-2}} \notag \\
&=\lim_{M\rightarrow\infty}\sqrt{1-2^{-2\mathbb{E}[u_k]}},
\end{align}
which implies that $ \sqrt{1-2^{-2\mathbb{E}[u_k]}} $ is also tight for $ \mathbb{E}\left[ \sqrt{1-2^{-2u_k}} \right] $ when $ M $ is large, namely $ \mathbb{E}\left[ \sqrt{1-2^{-2u_k}} \right] \approx \sqrt{1-2^{-2\mathbb{E}[u_k]}}, M \gg 1 $.

Recalling that $ U_k \approx U_k^\text{L} $, we finally have
\begin{align}
\label{secondterm}
\mathbb{E}\left[ \sqrt{1-2^{-2u_k}} \right]\approx\sqrt{1-2^{-2U_k^\text{L}}}, M \gg 1.
\end{align}
Substituting \eqref{firstterm} and \eqref{secondterm} into \eqref{Rtemp} yields \eqref{theorem1}.

\section{Proof of Theorem~\ref{theo2}}
\label{app:theo2}

We prove it for the case of the MF, stating that the proof of the ZF is similar to that of the MF, hence it is omitted here. For the MF precoder, we have
\ifCLASSOPTIONonecolumn
	\typeout{The onecolumn mode.}
	\begin{align}
	f_k^\text{MF}(\myvec{p}) &= -\sqrt{B^*}Q^{-1}(\epsilon_k)\log_2 \myexp \sqrt{1-\left( 1+\Gamma_k^\text{MF} \right)^{-2}}, \\
	\label{difff}
	\dfrac{\partial f_k^\text{MF}(\myvec{p})}{\partial p_k} &= -\dfrac{\sqrt{B^*}Q^{-1}(\epsilon_k)\log_2 \myexp \left( \dfrac{\Gamma_k^\text{MF}}{p_k}+\dfrac{(\Gamma_k^\text{MF})^2}{Mp_k} \right)}{\sqrt{1-\left( 1+\Gamma_k^\text{MF} \right)^{-2}}\left( 1+\Gamma_k^\text{MF} \right)^3}, \\
	\dfrac{\partial^2 f_k^\text{MF}(\myvec{p})}{\partial p_k^2} &= \dfrac{\sqrt{B^*}Q^{-1}(\epsilon_k)\log_2 \myexp \left( M+\Gamma_k^\text{MF} \right) \left( 3M-2+\Gamma_k^\text{MF}\right)}{\left( P_\text{B}-p_k+\varphi_k \right)^2\sqrt{1-\left( 1+\Gamma_k^\text{MF} \right)^{-2}}\left( 1+\Gamma_k^\text{MF} \right)^4}.
	\end{align}
\else
	\typeout{The twocolumn mode.}
	\begin{align}
	f_k^\text{MF}(\myvec{p}) &= -\sqrt{B^*}Q^{-1}(\epsilon_k)\log_2 \myexp \sqrt{1-\left( 1+\Gamma_k^\text{MF} \right)^{-2}}, \\
	\label{difff}
	\dfrac{\partial f_k^\text{MF}(\myvec{p})}{\partial p_k} &= -\dfrac{\sqrt{B^*}Q^{-1}(\epsilon_k)\log_2 \myexp \left( \dfrac{\Gamma_k^\text{MF}}{p_k}+\dfrac{(\Gamma_k^\text{MF})^2}{Mp_k} \right)}{\sqrt{1-\left( 1+\Gamma_k^\text{MF} \right)^{-2}}\left( 1+\Gamma_k^\text{MF} \right)^3},
	\end{align}
\fi
Since $ M \gg 1 \Rightarrow 3M-2 > 0  $, then $ f_k^\text{MF}(\myvec{p})<0 $, $ \partial f_k^\text{MF}(\myvec{p})/\partial p_k<0 $ and $ \partial^2 f_k^\text{MF}(\myvec{p})/\partial p_k^2>0 $, namely $ f_k^\text{MF}(\myvec{p}) $ is negative, differentiable, decreasing and convex.

\ifCLASSOPTIONtwocolumn
	\typeout{The twocolumn mode.}
	\begin{figure*}[!t]
		\centering
		\begin{align}
		\dfrac{\partial^2 f_k^\text{MF}(\myvec{p})}{\partial p_k^2} = \dfrac{\sqrt{B^*}Q^{-1}(\epsilon_k)\log_2 \myexp \left( M+\Gamma_k^\text{MF} \right) \left( 3M-2+\Gamma_k^\text{MF}\right)}{\left( P_\text{B}-p_k+\varphi_k \right)^2\sqrt{1-\left( 1+\Gamma_k^\text{MF} \right)^{-2}}\left( 1+\Gamma_k^\text{MF} \right)^4}.
		\end{align}
		\hrulefill
		\vspace*{-10pt}
	\end{figure*}
\fi

On the other hand, we have
\begin{align}
g_k^\text{MF}(\myvec{p}) &= B^* \log_2\left( 1+\Gamma_k^\text{MF} \right)-R_k (\Gamma_k,L_k,\epsilon_k), \\
\label{diffg}
\dfrac{\partial g_k^\text{MF}(\myvec{p})}{\partial p_k} &= \dfrac{B^*\left( \dfrac{\Gamma_k^\text{MF}}{p_k}+\dfrac{(\Gamma_k^\text{MF})^2}{Mp_k} \right)}{\left( 1+\Gamma_k^\text{MF} \right) \ln 2}, \\
\dfrac{\partial^2 g_k^\text{MF}(\myvec{p})}{\partial p_k^2} &= \dfrac{B^*\left( M+\Gamma_k^\text{MF} \right)\left( 2+\Gamma_k^\text{MF}-M \right)}{\left( P_\text{B}-p_k+\varphi_k \right)^2 \left( 1+\Gamma_k^\text{MF} \right)^2 \ln 2}.
\end{align}
Since $ M \gg 1 \Rightarrow 2+\Gamma_k^\text{MF}-M < 0  $, then $ g_k^\text{MF}(\myvec{p})>0 $, $ \partial g_k^\text{MF}(\myvec{p})/\partial p_k>0 $ and $ \partial^2 g_k^\text{MF}(\myvec{p})/\partial p_k^2<0 $, namely $ g_k^\text{MF}(\myvec{p}) $ is non-negative, differentiable, increasing and concave.

\begin{lemma}
\label{lem2}
Let $ \mathcal{C}\subseteq\mathbb{R}^n $ be a convex set and $ r: \mathcal{C}\rightarrow\mathbb{R} $. Then, $ r $ is QC if and only if its superlevel set $ \mathcal{S}_t=\left\lbrace \myvec{x} \in \mathcal{C} : r(\myvec{x}) \geqslant t \right\rbrace $ is convex for all $ t \in \mathbb{R} $.
\end{lemma}

\renewcommand{\IEEEQED}{\IEEEQEDopen}
\begin{IEEEproof}
The omitted proof can be found in~\cite{Boydbook,Zappone2015}.
\end{IEEEproof}
\renewcommand{\IEEEQED}{\IEEEQEDclosed}

\ifCLASSOPTIONtwocolumn
\typeout{The twocolumn mode.}
\begin{figure*}[!t]
	\centering
	\begin{align}
	\label{diff2h}
	& \dfrac{\partial^2 h_k^\text{MF}(\myvec{p})}{\partial p_k^2} = - \dfrac{\sqrt{B^*}\left( M+\Gamma_k^\text{MF} \right)}{\left( P_\text{B}-p_k+\varphi_k \right)^2 \sqrt{1-\left( 1+\Gamma_k^\text{MF} \right)^{-2}}\left( 1+\Gamma_k^\text{MF} \right)^4 \ln 2} \notag \\
	& \qquad \times \left[ Q^{-1}(\epsilon_k)\left( -3M+2-\Gamma_k^\text{MF} \right)-t\sqrt{B^*}\sqrt{1-\left( 1+\Gamma_k^\text{MF} \right)^{-2}}\left( 1+\Gamma_k^\text{MF} \right)^2\left( M-2-\Gamma_k^\text{MF}\right) \right].
	\end{align}
	\hrulefill
	\vspace*{-10pt}
\end{figure*}
\fi

\ifCLASSOPTIONonecolumn
	\typeout{The onecolumn mode.}
	According to Lemma~\ref{lem2}, the superlevel set $ \mathcal{S}_t=\left\lbrace \myvec{p} \in \mathcal{C} : f_k^\text{MF}(\myvec{p})/g_k^\text{MF}(\myvec{p}) \geqslant t \right\rbrace $ is taken into account. $ \mathcal{S}_t $ is the empty set for all $ t>0 $ when $ f_k^\text{MF}(\myvec{p})<0 $ and $ g_k^\text{MF}(\myvec{p})>0 $, thus only $ t \leqslant 0 $ has to be considered. The equivalent form of $ \mathcal{S}_t $ is given by $ \mathcal{S}_t=\left\lbrace \myvec{p} \in \mathcal{C} : f_k^\text{MF}(\myvec{p})-tg_k^\text{MF}(\myvec{p}) \geqslant 0 \right\rbrace $. Let $ h_k^\text{MF}(\myvec{p})=f_k^\text{MF}(\myvec{p})-tg_k^\text{MF}(\myvec{p}) $, then $ \partial^2 h_k^\text{MF}(\myvec{p})/\partial p_k^2=\partial^2 f_k^\text{MF}(\myvec{p})/\partial p_k^2-t \partial^2 g_k^\text{MF}(\myvec{p})/\partial p_k^2 $ is given by
	\begin{align}
	\label{diff2h}
	& \dfrac{\partial^2 h_k^\text{MF}(\myvec{p})}{\partial p_k^2} = - \dfrac{\sqrt{B^*}\left( M+\Gamma_k^\text{MF} \right)}{\left( P_\text{B}-p_k+\varphi_k \right)^2 \sqrt{1-\left( 1+\Gamma_k^\text{MF} \right)^{-2}}\left( 1+\Gamma_k^\text{MF} \right)^4 \ln 2} \notag \\
	& \qquad \times \left[ Q^{-1}(\epsilon_k)\left( -3M+2-\Gamma_k^\text{MF} \right)-t\sqrt{B^*}\sqrt{1-\left( 1+\Gamma_k^\text{MF} \right)^{-2}}\left( 1+\Gamma_k^\text{MF} \right)^2\left( M-2-\Gamma_k^\text{MF}\right) \right].
	\end{align}
	Because $ M \gg 1 $ and the order of magnitude on bandwidth is generally of kHz, we obtain
	\begin{align}
	&Q^{-1}(\epsilon_k)\left( -3M+2-\Gamma_k^\text{MF} \right) \notag \\
	&\quad - t\sqrt{B^*}\sqrt{1-\left( 1+\Gamma_k^\text{MF} \right)^{-2}}\left( 1+\Gamma_k^\text{MF} \right)^2\left( M-2-\Gamma_k^\text{MF}\right) \notag \\
	&\geqslant Q^{-1}(\epsilon_k)\left( -3M+2-\Gamma_k^\text{MF} \right) \notag \\
	&\quad - t\sqrt{B^*}\dfrac{\Gamma_k^\text{MF}}{1+\Gamma_k^\text{MF}}\left( M-2-\Gamma_k^\text{MF}\right) >0,
	\end{align}
	which states that the function $ h_k^\text{MF}(\myvec{p})=f_k^\text{MF}(\myvec{p})-tg_k^\text{MF}(\myvec{p}) $ is concave, namely the superlevel set $ \mathcal{S}_t $ is convex, and $ -\sqrt{L_k} $ is a QC function of $ p_k $.
\else
	\typeout{The twocolumn mode.}
	According to Lemma~\ref{lem2}, the superlevel set $ \mathcal{S}_t=\left\lbrace \myvec{p} \in \mathcal{C} : f_k^\text{MF}(\myvec{p})/g_k^\text{MF}(\myvec{p}) \geqslant t \right\rbrace $ is taken into account. $ \mathcal{S}_t $ is the empty set for all $ t>0 $ when $ f_k^\text{MF}(\myvec{p})<0 $ and $ g_k^\text{MF}(\myvec{p})>0 $, thus only $ t \leqslant 0 $ has to be considered. The equivalent form of $ \mathcal{S}_t $ is given by $ \mathcal{S}_t=\left\lbrace \myvec{p} \in \mathcal{C} : f_k^\text{MF}(\myvec{p})-tg_k^\text{MF}(\myvec{p}) \geqslant 0 \right\rbrace $. Let $ h_k^\text{MF}(\myvec{p})=f_k^\text{MF}(\myvec{p})-tg_k^\text{MF}(\myvec{p}) $, then $ \partial^2 h_k^\text{MF}(\myvec{p})/\partial p_k^2=\partial^2 f_k^\text{MF}(\myvec{p})/\partial p_k^2-t \partial^2 g_k^\text{MF}(\myvec{p})/\partial p_k^2 $ is given by \eqref{diff2h}. Because $ M \gg 1 $ and the order of magnitude on bandwidth is generally of kHz, we obtain
	\begin{align}
	&Q^{-1}(\epsilon_k)\left( -3M+2-\Gamma_k^\text{MF} \right) \notag \\
	&\quad - t\sqrt{B^*}\sqrt{1-\left( 1+\Gamma_k^\text{MF} \right)^{-2}}\left( 1+\Gamma_k^\text{MF} \right)^2\left( M-2-\Gamma_k^\text{MF}\right) \notag \\
	&\geqslant Q^{-1}(\epsilon_k)\left( -3M+2-\Gamma_k^\text{MF} \right) \notag \\
	&\quad - t\sqrt{B^*}\dfrac{\Gamma_k^\text{MF}}{1+\Gamma_k^\text{MF}}\left( M-2-\Gamma_k^\text{MF}\right) >0,
	\end{align}
	which states that the function $ h_k^\text{MF}(\myvec{p})=f_k^\text{MF}(\myvec{p})-tg_k^\text{MF}(\myvec{p}) $ is concave, namely the superlevel set $ \mathcal{S}_t $ is convex, and $ -\sqrt{L_k} $ is a QC function of $ p_k $.
\fi

\section{Proof of Lemma~\ref{lem1}}
\label{app:lem1}

Recall from \eqref{auxiliary} that
\begin{align}
F(\eta_{\tilde{\myvec{p}}}) &= \max_{\myvec{p}}\min_{k} \left\lbrace f_k(\myvec{p})-\min_{k}\left\lbrace \dfrac{f_k(\tilde{\myvec{p}})}{g_k(\tilde{\myvec{p}})} \right\rbrace g_k(\myvec{p}) \right\rbrace, \forall \myvec{p} \in \mathcal{P} \notag \\
&\overset{\text{(a)}}{\geqslant} \min_{k} \left\lbrace f_k(\tilde{\myvec{p}})- \min_{k}\left\lbrace \dfrac{f_k(\tilde{\myvec{p}})}{g_k(\tilde{\myvec{p}})} \right\rbrace g_k(\tilde{\myvec{p}}) \right\rbrace \notag \\
&\overset{\text{(b)}}{=} f_n(\tilde{\myvec{p}})- \dfrac{f_n(\tilde{\myvec{p}})}{g_n(\tilde{\myvec{p}})} g_n(\tilde{\myvec{p}}) = 0,
\end{align}
where the equality of (a) holds when $ \tilde{\myvec{p}} = \arg \max\limits_{\myvec{p}}\min\limits_{k} \left\lbrace f_k(\myvec{p})-\eta_{\tilde{\myvec{p}}} g_k(\myvec{p}) \right\rbrace $, and (b) follows from
\begin{align}
&\mathrel{\phantom{\Rightarrow}} \dfrac{f_n(\tilde{\myvec{p}})}{g_n(\tilde{\myvec{p}})} < \dfrac{f_k(\tilde{\myvec{p}})}{g_k(\tilde{\myvec{p}})}, \forall k \neq n, n=\arg\min_{k}\left\lbrace \dfrac{f_k(\tilde{\myvec{p}})}{g_k(\tilde{\myvec{p}})} \right\rbrace, \\
&\Rightarrow f_k(\tilde{\myvec{p}})-\dfrac{f_n(\tilde{\myvec{p}})}{g_n(\tilde{\myvec{p}})} g_k(\tilde{\myvec{p}}) > 0.
\end{align}

%
\bibliographystyle{IEEEtran}
\bibliography{IEEEabrv,Ref}

\begin{thebibliography}{10}
\providecommand{\url}[1]{#1}
\csname url@samestyle\endcsname
\providecommand{\newblock}{\relax}
\providecommand{\bibinfo}[2]{#2}
\providecommand{\BIBentrySTDinterwordspacing}{\spaceskip=0pt\relax}
\providecommand{\BIBentryALTinterwordstretchfactor}{4}
\providecommand{\BIBentryALTinterwordspacing}{\spaceskip=\fontdimen2\font plus
\BIBentryALTinterwordstretchfactor\fontdimen3\font minus
  \fontdimen4\font\relax}
\providecommand{\BIBforeignlanguage}[2]{{%
\expandafter\ifx\csname l@#1\endcsname\relax
\typeout{** WARNING: IEEEtran.bst: No hyphenation pattern has been}%
\typeout{** loaded for the language `#1'. Using the pattern for}%
\typeout{** the default language instead.}%
\else
\language=\csname l@#1\endcsname
\fi
#2}}
\providecommand{\BIBdecl}{\relax}
\BIBdecl

\bibitem{Qiu2018}
T.~{Qiu}, N.~{Chen}, K.~{Li}, M.~{Atiquzzaman}, and W.~{Zhao}, ``How can
  heterogeneous {Internet} of things build our future: {A} survey,''
  \emph{{IEEE} Commun. Surveys Tuts.}, vol.~20, no.~3, pp. 2011--2027,
  Thirdquarter 2018.

\bibitem{zhengsurvey2015}
K.~{Zheng}, Q.~{Zheng}, P.~{Chatzimisios}, W.~{Xiang}, and Y.~{Zhou},
  ``Heterogeneous vehicular networking: {A} survey on architecture, challenges,
  and solutions,'' \emph{{IEEE} Commun. Surveys Tuts.}, vol.~17, no.~4, pp.
  2377--2396, Fourthquarter 2015.

\bibitem{MacHardy2018}
Z.~{MacHardy}, A.~{Khan}, K.~{Obana}, and S.~{Iwashina}, ``{V2X} access
  technologies: {Regulation}, research, and remaining challenges,''
  \emph{{IEEE} Commun. Surveys Tuts.}, vol.~20, no.~3, pp. 1858--1877,
  Thirdquarter 2018.

\bibitem{adv2015}
K.~{Zheng}, Q.~{Zheng}, H.~{Yang}, L.~{Zhao}, L.~{Hou}, and P.~{Chatzimisios},
  ``Reliable and efficient autonomous driving: {T}he need for heterogeneous
  vehicular networks,'' \emph{{IEEE} Commun. Mag.}, vol.~53, no.~12, pp.
  72--79, Dec. 2015.

\bibitem{Wang2019}
J.~{Wang}, J.~{Liu}, and N.~{Kato}, ``Networking and communications in
  autonomous driving: {A} survey,'' \emph{{IEEE} Commun. Surveys Tuts.},
  vol.~21, no.~2, pp. 1243--1274, Secondquarter 2019.

\bibitem{Sun2016}
W.~{Sun}, E.~G. {Str\"om}, F.~{Br\"annstr\"om}, K.~C. {Sou}, and Y.~{Sui},
  ``Radio resource management for {D2D}-based {V2V} communication,''
  \emph{{IEEE} Trans. Veh. Technol.}, vol.~65, no.~8, pp. 6636--6650, Aug.
  2016.

\bibitem{Polyanskiy2010}
Y.~Polyanskiy, H.~V. Poor, and S.~Verdu, ``Channel coding rate in the finite
  blocklength regime,'' \emph{{IEEE} Trans. Inf. Theory}, vol.~56, no.~5, pp.
  2307--2359, May 2010.

\bibitem{Hayashi2009}
M.~{Hayashi}, ``Information spectrum approach to second-order coding rate in
  channel coding,'' \emph{{IEEE} Trans. Inf. Theory}, vol.~55, no.~11, pp.
  4947--4966, Nov. 2009.

\bibitem{Giuseppe2016}
G.~Durisi, T.~Koch, and P.~Popovski, ``Toward massive, ultrareliable, and
  low-latency wireless communication with short packets,'' \emph{Proc. {IEEE}},
  vol. 104, no.~9, pp. 1711--1726, Sep. 2016.

\bibitem{Yang2014}
W.~{Yang}, G.~{Durisi}, T.~{Koch}, and Y.~{Polyanskiy}, ``Quasi-static
  multiple-antenna fading channels at finite blocklength,'' \emph{{IEEE} Trans.
  Inf. Theory}, vol.~60, no.~7, pp. 4232--4265, Jul. 2014.

\bibitem{Zheng2016}
Q.~{Zheng}, K.~{Zheng}, H.~{Zhang}, and V.~C.~M. {Leung}, ``Delay-optimal
  virtualized radio resource scheduling in software-defined vehicular networks
  via stochastic learning,'' \emph{{IEEE} Trans. Veh. Technol.}, vol.~65,
  no.~10, pp. 7857--7867, Oct. 2016.

\bibitem{Zheng2015}
K.~{Zheng}, H.~{Meng}, P.~{Chatzimisios}, L.~{Lei}, and X.~{Shen}, ``An
  {SMDP}-based resource allocation in vehicular cloud computing systems,''
  \emph{{IEEE} Trans. Ind. Electron.}, vol.~62, no.~12, pp. 7920--7928, Dec.
  2015.

\bibitem{Yang2018}
G.~Yang, M.~Xiao, and H.~V. Poor, ``Low-latency millimeter-wave communications:
  {Traffic} dispersion or network densification?'' \emph{{IEEE} Trans.
  Commun.}, vol.~66, no.~8, pp. 3526--3539, Aug. 2018.

\bibitem{Forssell2019}
H.~{Forssell}, R.~{Thobaben}, H.~{Al-Zubaidy}, and J.~{Gross}, ``Physical layer
  authentication in mission-critical mtc networks: {A} security and delay
  performance analysis,'' \emph{{IEEE} J. Sel. Areas Commun.}, vol.~37, no.~4,
  pp. 795--808, Apr. 2019.

\bibitem{Choi2019}
J.~Choi, ``An effective capacity based approach to multi-channel low-latency
  wireless communications,'' \emph{{IEEE} Trans. Commun.}, vol.~67, no.~3, pp.
  2476--2486, Mar. 2019.

\bibitem{Yang2017}
H.~{Yang}, L.~{Zhao}, L.~{Lei}, and K.~{Zheng}, ``A two-stage allocation scheme
  for delay-sensitive services in dense vehicular networks,'' in \emph{Proc.
  IEEE International Conference on Communications Workshops (ICC Workshops)},
  Paris, France, May 2017, pp. 1358--1363.

\bibitem{Sebastian2015}
S.~Sebastian, G.~James, and A.-Z. Hussein, ``Delay analysis for wireless fading
  channels with finite blocklength channel coding,'' in \emph{Proc. ACM
  International Conference on Modeling, Analysis and Simulation of Wireless and
  Mobile Systems (MSWiM)}, Cancun, Mexico, Nov. 2015, pp. 13--22.

\bibitem{She2018}
C.~She, C.~Yang, and T.~Q.~S. Quek, ``Cross-layer optimization for
  ultra-reliable and low-latency radio access networks,'' \emph{{IEEE} Trans.
  Wireless Commun.}, vol.~17, no.~1, pp. 127--141, Jan. 2018.

\bibitem{Cui2012}
Y.~{Cui}, V.~K.~N. {Lau}, R.~{Wang}, H.~{Huang}, and S.~{Zhang}, ``A survey on
  delay-aware resource control for wireless systems--large deviation theory,
  stochastic {Lyapunov} drift, and distributed stochastic learning,''
  \emph{{IEEE} Trans. Inf. Theory}, vol.~58, no.~3, pp. 1677--1701, Mar. 2012.

\bibitem{Mei2018}
J.~{Mei}, K.~{Zheng}, L.~{Zhao}, Y.~{Teng}, and X.~{Wang}, ``A latency and
  reliability guaranteed resource allocation scheme for {LTE V2V} communication
  systems,'' \emph{{IEEE} Trans. Wireless Commun.}, vol.~17, no.~6, pp.
  3850--3860, Jun. 2018.

\bibitem{Johansson2015}
N.~A. {Johansson}, Y.~.~E. {Wang}, E.~{Eriksson}, and M.~{Hessler}, ``Radio
  access for ultra-reliable and low-latency {5G} communications,'' in
  \emph{Proc. IEEE International Conference on Communications Workshops (ICC
  Workshops)}, London, UK, Jun. 2015, pp. 1184--1189.

\bibitem{Marzetta2010}
T.~L. Marzetta, ``Noncooperative cellular wireless with unlimited numbers of
  base station antennas,'' \emph{{IEEE} Trans. Wireless Commun.}, vol.~9,
  no.~11, pp. 3590--3600, Nov. 2010.

\bibitem{zhengsurvey2015MIMO}
K.~{Zheng}, L.~{Zhao}, J.~{Mei}, B.~{Shao}, W.~{Xiang}, and L.~{Hanzo},
  ``Survey of large-scale {MIMO} systems,'' \emph{{IEEE} Commun. Surveys
  Tuts.}, vol.~17, no.~3, pp. 1738--1760, Thirdquarter 2015.

\bibitem{Zhao2019}
L.~{Zhao} and X.~{Wang}, ``Massive {MIMO} downlink for wireless information and
  energy transfer with energy harvesting receivers,'' \emph{{IEEE} Trans.
  Commun.}, vol.~67, no.~5, pp. 3309--3322, May 2019.

\bibitem{She2017}
C.~She, C.~Yang, and T.~Q.~S. Quek, ``Radio resource management for
  ultra-reliable and low-latency communications,'' \emph{{IEEE} Commun. Mag.},
  vol.~55, no.~6, pp. 72--78, Jun. 2017.

\bibitem{Hu2019}
Y.~{Hu}, Y.~{Zhu}, M.~C. {Gursoy}, and A.~{Schmeink}, ``{SWIPT}-enabled
  relaying in {IoT} networks operating with finite blocklength codes,''
  \emph{{IEEE} J. Sel. Areas Commun.}, vol.~37, no.~1, pp. 74--88, Jan. 2019.

\bibitem{Sun2019}
C.~{Sun}, C.~{She}, C.~{Yang}, T.~Q.~S. {Quek}, Y.~{Li}, and B.~{Vucetic},
  ``Optimizing resource allocation in the short blocklength regime for
  ultra-reliable and low-latency communications,'' \emph{{IEEE} Trans. Wireless
  Commun.}, vol.~18, no.~1, pp. 402--415, Jan. 2019.

\bibitem{Zhou2019}
L.~{Zhou}, A.~{Wolf}, and M.~{Motani}, ``On lossy multi-connectivity: Finite
  blocklength performance and second-order asymptotics,'' \emph{{IEEE} J. Sel.
  Areas Commun.}, vol.~37, no.~4, pp. 735--748, Apr. 2019.

\bibitem{Lv2015}
Y.~{Lv}, Y.~{Duan}, W.~{Kang}, Z.~{Li}, and F.~{Wang}, ``Traffic flow
  prediction with big data: {A} deep learning approach,'' \emph{{IEEE} Trans.
  Intell. Transp. Syst.}, vol.~16, no.~2, pp. 865--873, Apr. 2015.

\bibitem{Polson2017}
N.~G. Polson and V.~O. Sokolov, ``Deep learning for short-term traffic flow
  prediction,'' \emph{Transp. Res. Part C Emerg. Technol.}, vol.~79, pp. 1--17,
  Jun. 2017.

\bibitem{trafficflow}
P.~Kachroo and K.~M.~A. {\"O}zbay, ``Traffic flow theory,'' in \emph{Feedback
  Control Theory for Dynamic Traffic Assignment}.\hskip 1em plus 0.5em minus
  0.4em\relax Cham: Springer, 2018, pp. 57--87.

\bibitem{Kashyap2016}
S.~{Kashyap}, E.~{Bj\"ornson}, and E.~G. {Larsson}, ``On the feasibility of
  wireless energy transfer using massive antenna arrays,'' \emph{{IEEE} Trans.
  Wireless Commun.}, vol.~15, no.~5, pp. 3466--3480, May 2016.

\bibitem{Zhao2017}
L.~{Zhao}, T.~{Riihonen}, W.~{Xiang}, Y.~{Kuang}, and K.~{Zheng}, ``Resource
  optimization of wireless information and energy supply control systems with
  massive {MIMO},'' \emph{{IEEE} Commun. Lett.}, vol.~21, no.~12, pp.
  2734--2737, Dec. 2017.

\bibitem{Zhang2018}
J.~{Zhang}, L.~{Dai}, X.~{Li}, Y.~{Liu}, and L.~{Hanzo}, ``On low-resolution
  {ADCs} in practical {5G} millimeter-wave massive {MIMO} systems,''
  \emph{{IEEE} Commun. Mag.}, vol.~56, no.~7, pp. 205--211, Jul. 2018.

\bibitem{Ngo2013}
H.~Q. Ngo, E.~G. Larsson, and T.~L. Marzetta, ``Energy and spectral efficiency
  of very large multiuser {MIMO} systems,'' \emph{{IEEE} Trans. Commun.},
  vol.~61, no.~4, pp. 1436--1449, Apr. 2013.

\bibitem{Liu2018}
X.~{Liu}, Y.~{Li}, L.~{Xiao}, and J.~{Wang}, ``Performance analysis and power
  control for multi-antenna {V2V} underlay massive {MIMO},'' \emph{{IEEE}
  Trans. Wireless Commun.}, vol.~17, no.~7, pp. 4374--4387, Jul. 2018.

\bibitem{3gpp38211}
``{NR} physical channels and modulation,'' 3GPP, Tech. Rep. TS 38.211 V15.6.0,
  Jun. 2019.

\bibitem{Abbas2019}
F.~{Abbas}, P.~{Fan}, and Z.~{Khan}, ``A novel low-latency {V2V} resource
  allocation scheme based on cellular {V2X} communications,'' \emph{{IEEE}
  Trans. Intell. Transp. Syst.}, vol.~20, no.~6, pp. 2185--2197, Jun. 2019.

\bibitem{Park2019}
Y.~{Park}, T.~{Kim}, and D.~{Hong}, ``Resource size control for reliability
  improvement in cellular-based {V2V} communication,'' \emph{{IEEE} Trans. Veh.
  Technol.}, vol.~68, no.~1, pp. 379--392, Jan. 2019.

\bibitem{Fleury1996}
B.~H. Fleury, ``An uncertainty relation for {WSS} processes and its application
  to {WSSUS} systems,'' \emph{{IEEE} Commun. Mag.}, vol.~44, no.~12, pp.
  1632--1634, Dec. 1996.

\bibitem{Dahlmanbook}
E.~Dahlman, S.~Parkvall, and J.~Skold, \emph{4G: LTE/LTE-Advanced for Mobile
  Broadband}.\hskip 1em plus 0.5em minus 0.4em\relax Cambridge, Massachusetts:
  Academic Press, 2013.

\bibitem{Shen2018}
K.~{Shen} and W.~{Yu}, ``Fractional programming for communication systems-{Part
  I}: {Power} control and beamforming,'' \emph{{IEEE} Trans. Signal Process.},
  vol.~66, no.~10, pp. 2616--2630, May 2018.

\bibitem{Zappone2015}
A.~Zappone and E.~Jorswieck, ``Energy efficiency in wireless networks via
  fractional programming theory,'' \emph{Foundations and Trends in
  Communications and Information Theory}, vol.~11, no. 3-4, pp. 185--396, Jun.
  2015.

\bibitem{Borweinbook}
J.~M. Borwein and P.~B. Borwein, \emph{{Pi} and the {AGM}: {A} Study in
  Analytic Number Theory and Computational Complexity}.\hskip 1em plus 0.5em
  minus 0.4em\relax Hoboken, New Jersey: John Wiley \& Sons, 1987.

\bibitem{Cheung2013}
K.~T.~K. {Cheung}, S.~{Yang}, and L.~{Hanzo}, ``Achieving maximum
  energy-efficiency in multi-relay {OFDMA} cellular networks: {A} fractional
  programming approach,'' \emph{{IEEE} Trans. Commun.}, vol.~61, no.~7, pp.
  2746--2757, Jul. 2013.

\bibitem{Boydbook}
S.~Boyd and L.~Vandenberghe, \emph{Convex Optimization}.\hskip 1em plus 0.5em
  minus 0.4em\relax Cambridge, England: Cambridge University Press, 2004.

\end{thebibliography}
\end{document}